\documentclass[english,aps,prb]{revtex4}

\usepackage[T1]{fontenc}
\usepackage[latin9]{inputenc}
\usepackage{float}
\usepackage{amsmath}
\usepackage{graphicx}
\usepackage{setspace}
\usepackage{amssymb}
\usepackage{url}
\usepackage[breaklinks=false]{hyperref} 
\usepackage{indentfirst} 
\usepackage{subfigure}
\usepackage{footmisc} 



\usepackage{babel}

\begin{document}

\title{Quasiparticle scattering from vortices in d-wave superconductors II : Berry phase contribution}

\author{Sriram Ganeshan$^{1}$, Manas Kulkarni$^{1,2}$, Adam C. Durst$^{1}$}

\affiliation{$^{1}$Department of Physics and Astronomy, Stony Brook University,
Stony Brook, NY 11794-3800}

\affiliation{$^{2}$Department of Condensed Matter Physics and Material Science,
Brookhaven National Laboratory, Upton, NY-11973}

\maketitle
In the mixed state of a d-wave superconductor, Bogoliubov quasiparticles are scattered from magnetic vortices
via a combination of two effects: Aharonov-Bohm scattering due to the Berry phase acquired by a quasiparticle
upon circling a vortex, and effective potential scattering due to the superflow swirling about the vortices. In
this paper, we consider the Berry phase contribution in the absence of superflow, which results in branch cuts
between neighboring vortices across which the quasiparticle wave function changes sign. Here, the simplest
problem that captures the physics is that of scattering from a single finite branch cut that stretches between two
vortices. Elliptical coordinates are natural for this two-center problem, and we proceed by separating the massless
Dirac equation in elliptical coordinates. The separated equations take the form of the Whittaker-Hill equations,
which we solve to obtain radial and angular eigenfunctions. With these eigenfunctions in hand, we construct the
scattering cross section via partial wave analysis. We discuss the scattering effect of Berry phase in the absence
of superflow, having considered the superflow effect in the absence of Berry phase in a separate paper. We also
provide qualitative comparison of transport cross sections for the Berry phase and the Superflow effects. The
important issue of interference between the two effects is left to future work.
\maketitle


\section{Introduction}
 Massless Dirac quasiparticles are the low energy excitations of a $d$-wave superconductor. In the vortex state, low temperature transport properties, such as longitudinal thermal conductivity and thermal Hall conductivity, can be explained by studying the scattering of these quasiparticles from magnetic vortices. Quasiparticles scatter from vortices via two basic mechanisms: a circulating superflow and a Berry phase factor of (-1) acquired upon circling a vortex. Scattering due to superflow current in a single vortex (without the Berry phase effect) has been calculated in Refs.~\onlinecite{adamprl} and \onlinecite{superflow1}. Although the transport properties calculated in Ref.~\onlinecite{adamprl} capture the essential physics qualitatively, the Berry phase effect becomes important for the higher field (lower temperature) regime where the deBroglie wavelength is comparable to the distance between vortices. It is therefore of interest to consider the effect of Berry phase on the quasiparticle scattering. The influence of the Aharonov-Bohm effect due to an isolated vortex line on the quasiparticle states has been studied in the Refs \onlinecite{melnikov} and \onlinecite{jiye}. Ref  \onlinecite{melnikov} also obtains the quasiparticle wavefunction and density of states at distances from the vortex large compared to the pentration depth ($r >> \lambda$). In this paper, we consider ($r << \lambda$) and work in the limit  $\lambda \rightarrow \infty$, such that the magnetic field is constant across the sample. This limit permits the application of a singular gauge transformation that encodes the Berry phase effect in the form of anti-periodic boundary conditions on the quasiparticle wave function. In contrast to conventional superconductors, the density of states at low energies in d-wave systems is dominated by contributions which come from the regions far from the cores\cite{volovik_1993} and are associated with extended quasiparticle states with momenta close to the nodal directions. This allows us to neglect the vortex core physics by setting the coherence length ($\xi \rightarrow 0$). This conclusion, based on the semiclassical approach \cite{jmelnikov, melnikov}, has been confirmed by numerical analysis \cite{franztesano} of the BdG equations for a single isolated vortex line (in the limit $\lambda \rightarrow \infty$). Note that the calculations presented in \onlinecite{franztesano} also point to the absence of truly localized core states or any resonant levels in the pure d-wave case, though such states were observed in numerical simulations \cite{morita}. AB effect on quasiparticle excitations in macroscopic superconducting rings has been studied in detail in Refs \onlinecite{barash} and \onlinecite{juricic}. In this work, we calculate the scattering contribution due to the Berry phase effect. We take the following path. As discussed in Ref.~\onlinecite{superflow1} and summarized in Appendix~\ref{sec:BdG}, we apply a singular gauge transformation to the Bogoliubov-de Gennes equation and shift the origin of momentum space to the location of one of the gap nodes. This reduces the problem to that of an (anisotropic) Dirac fermion scattering from an effective non-central potential (due to the superflow) in the presence of antiperiodic boundary conditions (a consequence of our gauge choice). For simplicity, we neglect the anisotropy of the Dirac dispersion by taking $v_{f}=v_{2}$. Since we are only interested in the Berry phase contribution, we neglect the presence of the effective non-central potential, which further reduces the problem to that of a massless Dirac fermion scattering due to the antiperiodic boundary conditions. Within the single vortex approximation, the antiperiodic boundary condition manifests as a semi-infinite branch cut terminating at the vortex core. In Sec.~\ref{app:vison} we study the single vortex scattering of quasiparticles due to this semi-infinite branch cut (without superflow current) and obtain a divergent differential cross section in the forward direction. The divergent nature of this cross section is unphysical and requires that we regularize this semi-infinite branch cut. In real situations these branch cuts terminate on the cores of the neighboring vortices and are finite in nature. Thus, considering a pair of vortices as our scatterer captures the finite branch cut and regularizes the Berry phase effect. Elliptical coordinates serve as a natural choice for this problem with two vortex cores. The presence of a finite branch cut between the two vortices manifests itself as a boundary condition on the wave function spinor across the line segment joining the two foci of the ellipse. In elliptical coordinates $(\mu,\nu)$, we simply write  \begin{equation} \psi(\mu,-\nu)\mid_{\mu=0}=(-1)^B\ \psi(\mu,\nu)\mid_{\mu=0}\\ \label{bound}\end{equation} We can impose the Berry phase condition using parameter B ($B=0,1$). When B=0, there is no branch cut between the vortex cores. We can turn on the Berry phase (branch cut) between the vortex cores by setting B=1. Thus, neglecting the anisotropy of the gap nodes, our problem reduces to that of the scattering of massless Dirac quasiparticles due to this finite branch cut. A similar setup for the scattering of quasiparticles due to the Berry phase has been considered by Melikyan and Tesanovic in Ref.~\onlinecite{tesanovic}. Their approach was to construct scattering solutions to the 2d Dirac equation in elliptical coordinates from solutions to the Klein-Gordon equation (separated in elliptical coordinates) using self-adjoint extensions. But in going from Klein-Gordon to Dirac, the solutions are not separated in elliptical coordinates anymore. This results in not being able to construct all the self-adjoint extensions. In this paper, we avoid the issue of self-adjoint extensions by taking a different approach. We directly separate the (2+1)d Dirac equation in elliptical coordinates\cite{1990-Villalba,1982-cook}. The separation of variables is done in Sec.~\ref{sec:Separation-of-Dirac}. The separated equations are Whittaker Hill equations (WHE)\cite{1990-Villalba, 1991-Villalba}. In Sec.~\ref{sec:Solutions-to-Whittakker-Hill's} we solve the separated equations to obtain eigenfunctions\cite{urwin,arscott1967,arscottbook, figuerdo,figuerdo2,figuerdo3,leaver,ince,Ronveaux}. In Sec.~\ref{sec:Expansion-of-plane}, we develop an expansion for the incident plane wave (representing quasiparticle current) in terms of the separated solutions of WHE. In Sec.~\ref{sec-construction} we construct the scattering amplitude (as a sum of partial waves) from the asymptotic form of the scattered wave. We impose the boundary condition Eq.~(\ref{bound}) on the full wavefunction spinor and calculate the phase shifts for each partial wave. We show that for B=0 (no branch cut) there is no quasiparticle scattering. In Sec~\ref{sec:Scattering-cross-section}, we turn on the Berry phase by setting B=1 (with branch cut between cores) and obtain a non-zero scattering cross section. Results and analysis of the quasiparticle scattering cross section in the presence of Berry phase is presented in Sec.~\ref{results}. Conclusions are discussed in Sec.~\ref{sec:conclude}.

\section{ Berry Phase Scattering of Incident Plane Wave in single vortex approximation (without superflow)}
\label{app:vison}

Quasiparticles scatter from vortices via both the circulating superflow and the Berry phase factor of (-1) acquired upon circling the vortex. This phase is encoded in the antiperiodic boundary conditions imposed on quasiparticles in our chosen gauge (see Appendix \ref{sec:BdG} for details). In this section, we neglect the superflow by setting $P_{s}=0$ and consider only the Berry phase contribution. That is, we consider the scattering of quasiparticles due only to the presence of antiperiodic boundary conditions. Furthermore, we neglect the anisotropy of the Dirac dispersion and take $v_{f}=v_{2}$ ($\alpha=1$). As we shall see, the antiperiodic boundary conditions yield the Aharonov-Bohm interference effect of an enclosed $\pi$-flux.

We consider the isotropic Dirac Hamiltonian
\begin{equation}
H = v_{f} \left[ \tau_{3} p_{x} + \tau_{1} p_{y} \right]
\label{eq:HD0}
\end{equation}
and seek solutions to the Bogoliubov-de Gennes equation, $H\Phi=E\Phi$. We express the quasiparticle wave function as a linear combination of angular momentum eigenstates which satisfy $J\Phi_{\mu}=\mu\Phi_{\mu}$. Since we have neglected the superflow, there is no effective potential and the general solution is easily found to be
\begin{equation}
\Phi = \sum_{\mu} \left[
(A_{\mu} J_{\mu-\frac{1}{2}} + B_{\mu} Y_{\mu-\frac{1}{2}})
e^{i(\mu-\frac{1}{2})\phi}
\left( \begin{array}{c} 1 \\ i \end{array} \right)
+ i(A_{\mu} J_{\mu+\frac{1}{2}} + B_{\mu} Y_{\mu+\frac{1}{2}})
e^{i(\mu+\frac{1}{2})\phi}
\left( \begin{array}{c} 1 \\ -i \end{array} \right) \right]
\label{eq:gensol}
\end{equation}
where $A_{\mu}$ and $B_{\mu}$ are complex constants and $J_{\mu \pm \frac{1}{2}}$ and $Y_{\mu \pm \frac{1}{2}}$ are
Bessel functions of argument $\rho=kr$. However, rather than imposing periodic boundary conditions by requiring that
$\mu=n+1/2$ with $n=\mbox{integer}$, here we shall impose antiperiodic boundary conditions,
\begin{equation}
\Phi(r,\phi+2\pi) = -\Phi(r,\phi)
\label{eq:antiper}
\end{equation}
by requiring that $\mu=\mbox{integer}$. The radial functions are therefore half-integer Bessel functions rather than integer Bessel functions. The coefficients, $A_{\mu}$ and $B_{\mu}$, are determined by satisfying boundary conditions at both long and short distances.

At long distances, we require an asymptotic wave function that is equal to the sum of an incident plane wave, $\Phi_{i}$, and an outgoing radial wave, $\Phi_{s}$. In terms of the current functional discussed in Appendix \ref{sec:BdG}, ${\bf j}[\Phi] = v_{f} \Phi^{\dagger} (\tau_{3} \hat{\bf x} + \tau_{1} \hat{\bf y}) \Phi$, we require that ${\bf j}[\Phi_{i}] \sim \hat{\bf k}$ and ${\bf j}[\Phi_{s}] \sim \hat{\bf r}$. In the presence of antiperiodic boundary conditions, we seek an incident wave of the form
\begin{equation}
\Phi_{i}({\bf r}) = e^{i\gamma\frac{\varphi}{2}}
e^{i{\bf k} \cdot {\bf r}}
\left( \begin{array}{c} \cos \frac{\theta}{2} \\
\sin \frac{\theta}{2} \end{array} \right)
\;\;\;\;\;\;\;\;
\gamma = \pm 1
\label{eq:incid}
\end{equation}
and a scattered wave of the form
\begin{equation}
\Phi_{s}({\bf r}) = f(\varphi) \frac{e^{ikr}}{\sqrt{r}}
\left( \begin{array}{c} \cos \frac{\phi}{2} \\
\sin \frac{\phi}{2} \end{array} \right)
\label{eq:scat}
\end{equation}
where ${\bf k}=(k,\theta)$, ${\bf r}=(r,\phi)$, and $\varphi=\phi-\theta$. In analogy with the problem of Aharonov-Bohm \cite{aharnov} scattering from an enclosed magnetic flux, we can say that $\gamma=-1$ corresponds to an effective $\pi$-flux while $\gamma=+1$ corresponds to an effective $(-\pi)$-flux. Since these two cases are equivalent, the choice of $\gamma=\pm 1$ is arbitrary. In the asymptotic limit, the half-integer Bessel functions take the form
\begin{equation}
J_{\mu-\frac{1}{2}}(\rho) = \eta_{\mu} \sqrt{\frac{2}{\pi\rho}} \cos
(\rho - |\mu|\pi/2) \;\;\;\;\;\;\;\;
Y_{\mu-\frac{1}{2}}(\rho) = \eta_{\mu} \sqrt{\frac{2}{\pi\rho}} \sin
(\rho - |\mu|\pi/2)
\label{eq:halfint}
\end{equation}
where $\eta_{\mu}=1$ for $\mu > 0$ and $\eta_{\mu}=(-1)^{\mu}$ for $\mu \leq 0$. Proceeding along the lines of Ref.~\onlinecite{superflow1}, we can plug these asymptotic expressions into Eq.~(\ref{eq:gensol}), reorganize terms, and thereby obtain a suggestive (yet still general) form for the quasiparticle wave function. Doing so, we find that
\begin{equation}
\Phi = e^{i\gamma\frac{\varphi}{2}} e^{i{\bf k} \cdot {\bf r}}
\left(\begin{array}{c} \cos\frac{\theta}{2} \\
\sin\frac{\theta}{2} \end{array}\right)
+ f(\varphi) \frac{e^{ikr}}{\sqrt{r}}
\left(\begin{array}{c} \cos\frac{\phi}{2} \\
\sin\frac{\phi}{2} \end{array}\right)
- i \gamma g(\varphi) \frac{e^{-ikr}}{\sqrt{r}}
\left(\begin{array}{c} -\sin\frac{\phi}{2} \\
\cos\frac{\phi}{2} \end{array}\right)
\label{eq:reorgsol}
\end{equation}
where
\begin{equation}
f(\varphi) \equiv \sqrt{\frac{2}{\pi k}} \sum_{\mu} b_{\mu} e^{i\mu\varphi}
\;\;\;\;\;\;\;\;
g(\varphi) \equiv \sqrt{\frac{2}{\pi k}} \sum_{\mu} a_{\mu} e^{i\mu\varphi}
\label{eq:visfgphi}
\end{equation}
and $a_{\mu}$ and $b_{\mu}$ are complex constants defined via
\begin{equation}
A_{\mu}-iB_{\mu} \equiv i^{\mu} e^{-i\mu\theta}
\left( e^{-i\frac{\pi}{4}}/2 + b_{\mu} \right)
\label{eq:ABbmu}
\end{equation}
\begin{equation}
A_{\mu}+iB_{\mu} \equiv -\gamma i^{\mu} e^{-i\mu\theta}
\left( e^{i\frac{\pi}{4}}/2 + (-1)^{\mu} a_{\mu} \right) .
\label{eq:ABamu}
\end{equation}
If the plane wave is to be the only incident wave, we must eliminate the incident radial wave by requiring that $a_{\mu}=0$ for all $\mu$. With this restriction, $B_{\mu}$ and $b_{\mu}$ are related to $A_{\mu}$ via
\begin{equation}
B_{\mu} = i \left( A_{\mu} + i\gamma \ A_{\mu}^{0} \right)
\;\;\;\;\;\;\;\;
b_{\mu} = e^{-i\frac{\pi}{4}} \left( \frac{A_{\mu}}{A_{\mu}^{0}}
-\frac{1-i\gamma}{2} \right)
\label{eq:BmubmufromAmu}
\end{equation}
where $A_{\mu}^{0} \equiv i^{\mu-1/2} e^{-i\mu\theta}/2$.
The asymptotic wave function then takes the desired form
\begin{equation}
\Phi = e^{i\gamma\frac{\varphi}{2}} e^{i{\bf k} \cdot {\bf r}}
\left(\begin{array}{c} \cos\frac{\theta}{2} \\
\sin\frac{\theta}{2} \end{array}\right)
+ f(\varphi) \frac{e^{ikr}}{\sqrt{r}}
\left(\begin{array}{c} \cos\frac{\phi}{2} \\
\sin\frac{\phi}{2} \end{array}\right)
\label{eq:asympsol}
\end{equation}
and the differential cross section is given by
\begin{equation}
\frac{d\sigma}{d\varphi} = |f(\varphi)|^{2}
= \frac{2}{\pi k} \left| \sum_{\mu} b_{\mu} e^{i\mu\varphi} \right|^{2} .
\label{eq:visdCSdef}
\end{equation}

We can now determine the $b_{\mu}$ by imposing appropriate boundary conditions at the origin. As discussed in Ref.~\onlinecite{superflow1}, the most restrictive condition is that the current through the origin (a point of zero area) must be zero. More precisely, we consider a semicircle of radius $\epsilon$, oriented about the $\hat{\bf \theta}$ direction, and require that the total current passing through it, $I_{\theta}$, vanishes as $\epsilon \rightarrow 0$.
If $\Phi(\rho \rightarrow 0) \sim \rho^{\alpha}$, then $I_{\theta} \sim \epsilon^{2\alpha+1}$. Thus, to ensure that
$I_{\theta}$ does not diverge at the origin, we must eliminate all terms in Eq.~(\ref{eq:gensol}) which diverge faster than $\rho^{-1/2}$ as $\rho \rightarrow 0$. Since the half-integer Bessel functions exhibit the small-argument behavior, $J_{\mu\pm\frac{1}{2}} \sim \rho^{\mu\pm\frac{1}{2}}$ and $Y_{\mu\pm\frac{1}{2}} \sim \rho^{-\mu\mp\frac{1}{2}}$, this clearly requires that
\begin{equation}
\mbox{$B_{\mu}=0$ for $\mu>0$}
\;\;\;\;\;\;\;\;
\mbox{$A_{\mu}=0$ for $\mu<0$}
\label{eq:condBA}
\end{equation}
The condition for $\mu=0$ is more subtle. Enforcing the above, the resulting wave function is dominated, as $\rho \rightarrow 0$, by the terms which diverge exactly as $\rho^{-1/2}$. We therefore find that
\begin{equation}
\Phi(\rho \rightarrow 0) = \sqrt{\frac{2}{\pi\rho}}
\left[ (A_{0}-iB_{0}) \left(\begin{array}{c} \cos\frac{\theta}{2} \\
\sin\frac{\theta}{2} \end{array}\right)
+ i (A_{0}+iB_{0}) \left(\begin{array}{c} -\sin\frac{\theta}{2} \\
\cos\frac{\theta}{2} \end{array}\right) \right] .
\label{eq:Phi0}
\end{equation}
From Eq.~(\ref{eq:BmubmufromAmu}), we know that $A_{0}+iB_{0}=-i\gamma A_{0}^{0}$ where $A_{0}^{0}=e^{-i\pi/4}/2$.
Furthermore, we can define a complex constant, $\beta$, such that $A_{0}-iB_{0} \equiv -i\gamma A_{0}^{0} \beta$. With this definition, the current density near the origin takes the form
\begin{equation}
{\bf j}(\rho \rightarrow 0) = \frac{v_{f}}{2\pi\rho}
\left[ \left( |\beta|^{2} - 1 \right) \hat{r}
+ 2 \,\mbox{Im}[\beta] \hat{\phi} \right] .
\label{eq:currden0}
\end{equation}
Explicitly computing the current through the origin, we find that
\begin{equation}
I_{\theta} = \lim_{\epsilon \rightarrow 0}
\int_{\theta-\pi/2}^{\theta+\pi/2} \!\epsilon \,d\phi\,
{\bf j}(\epsilon) \cdot \hat{\bf \theta}
= v_{f} \int \frac{d\phi}{2\pi}
\left[ \left( |\beta|^{2} - 1 \right) \hat{r} \cdot \hat{\theta}
+ 2 \,\mbox{Im}[\beta] \hat{\phi} \cdot \hat{\theta} \right]
\label{eq:Itheta0}
\end{equation}
which must be set to zero for all directions $\hat{\bf \theta}$. This requires that $\beta=\pm 1$.

Putting everything together yields the values of our original coefficients
\begin{equation}
A_{\mu} = -i\gamma A_{\mu}^{0} \left\{
\begin{array}{cc} 1 & \mu>0 \\
\frac{1+\beta}{2} & \mu=0 \\
0 & \mu<0 \end{array} \right\}
\;\;\;\;\;\;\;\;
B_{\mu} = -\gamma A_{\mu}^{0} \left\{
\begin{array}{cc} 0 & \mu>0 \\
\frac{1-\beta}{2} & \mu=0 \\
1 & \mu<0 \end{array} \right\}
\label{eq:AmuBmu}
\end{equation}
where $\gamma=\pm 1$ and $\beta=\pm 1$. The $Z_{2}$ ambiguity in $\gamma$ and $\beta$ is a consequence of the equivalence of a $\pi$-flux with a $(-\pi)$-flux, which cannot affect observable quantities. For $\beta=\gamma=\pm 1$,
\begin{equation}
b_{\mu}=\frac{1}{\sqrt{2}} \left\{
\begin{array}{cc} i & \gamma\mu > 0 \\ -1 & \gamma\mu \leq 0 \end{array}
\right\} .
\label{eq:bmuresult}
\end{equation}
Plugging this into Eq.~(\ref{eq:visdCSdef}) and summing over $\mu$ (with a convergence factor $e^{-|\mu| 0^{+}}$), yields
\begin{equation}
\frac{d\sigma}{d\varphi} = \frac{1}{2\pi k\,\sin^{2}(\varphi/2)}
- \gamma \frac{2}{k} \frac{\delta(\varphi)}{\varphi} .
\label{eq:visdCSdelta}
\end{equation}
The same result is obtained for $\beta=-\gamma$. Note, however, that the above is only valid for $\varphi \neq 0$.
As discussed (for the electron scattering case) in the original paper by Aharonov and Bohm \cite{aharnov}, as well as in an excellent review by Olariu and Popescu \cite{popescu}, our asymptotic approximations are only valid away from the forward direction. Thus, the second term above, which is only nonzero for $\varphi=0$ and is an artifact of our casual
treatment of the forward direction, can be dropped. (For a detailed treatment of the Aharonov-Bohm scattering of an electron in the forward direction, see the paper by Stelitano \cite{stelitano} Our differential cross section therefore takes the form
\begin{equation}
\frac{d\sigma}{d\varphi} = \frac{1}{2\pi k\,\sin^{2}(\varphi/2)}
\label{eq:visdCS}
\end{equation}
which is exactly the differential cross section for the Aharonov-Bohm scattering of an electron from an enclosed $\pi$-flux. As expected, this result is independent of our choice of $\gamma=\pm 1$ and $\beta=\pm 1$ and is the same for quasiparticles about any of the four gap nodes. Due to the infinite range of the Berry phase effect, the total
cross section diverges. However, the transport cross section is finite and given by $\sigma_{\parallel} = 1/\pi k$.
Since left-right symmetry is not broken in the absence of a superflow, the skew cross section is zero. In the zero-superflow case considered above, it was easy enough to neglect the subtleties associated with forward scattering in the presence of antiperiodic boundary conditions. However, if we were to consider the superflow and the Berry phase effects together, it would be necessary to treat such nuances more carefully. The first step towards that is to regularize the calculation of the cross section due to the Berry phase effect. This is the goal of the remainder of this paper.

\section{Regularization of Berry phase in double vortex setup}
\label{sec:regularize}

\begin{figure}[H]
\noindent \begin{center}
\includegraphics[scale=0.5]{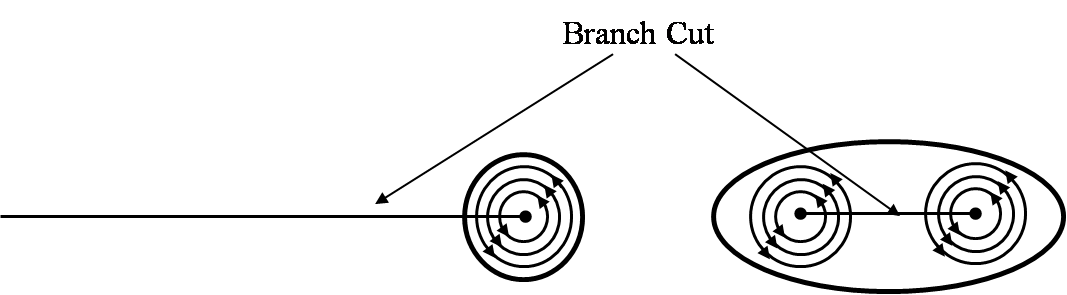}
\par\end{center}%
\caption{Single vortex with semi-infinite branch cut and double vortex with finite branch cut due to the Berry phase }
\label{fig:svbc}
\end{figure}

The infinite range of the Berry phase effect is solely due to the isolated treatment of the single vortex (see Fig.~\ref{fig:svbc}). In reality these vortices are not isolated and the Berry phase effect terminates at the neighboring vortices thereby resulting in a finite branch cut. The simplest object containing a finite branch cut is a pair of vortices separated by some distance as shown in Fig.~\ref{fig:svbc}. Elliptical coordinates is a natural setting for studying two-center problems. We define elliptical coordinates via
\begin{eqnarray}
x & = & R\,\cosh\mu\,\cos\nu\\
y & = & R\,\sinh\mu\,\sin\nu\end{eqnarray}
The presence of a finite branch cut between the two vortex cores can be imposed as a boundary condition at $\mu=0$ (line separating foci) on the full wavefunction spinor in the following way \begin{equation} \psi(\mu,-\nu)\mid_{\mu=0}=(-1)^B\ \psi(\mu,\nu)\mid_{\mu=0}\\\label{bcond}\end{equation}
Parameter B is defined to take values 0 or 1. For B=0, there is no branch cut between the pair of vortices (or foci) and the condition reduces to a trivial continuity condition at $\mu=0$ of the elliptical coordinate system. Setting B=1, we turn on the Berry phase effect via non-trivial boundary condition. The presence of the branch cut captures the fact that when quasiparticle passes between the vortex cores it acquires a phase factor of (-1), which on its own can scatter quasiparticles.
Our task for the rest of this paper is to capture the scattering of quasiparticles due to a finite branch cut between two vortices. We will follow the same prescription that we applied to the single vortex case to calculate the scattering cross section. The first step is to separate the (2+1)d Dirac equation in elliptical coordinates. The second step is to solve the separated equations to get the eigenfunctions for the separation parameter. Third step is to construct incoming plane wave and outgoing scattered wave in terms of phase shifts from the separated eigenfunctions. Fourth and final step is to impose the boundary condition for the branch cut on the full wavefunction spinor and calculate the scattering
amplitude and differential scattering cross section.

\section{\label{sec:Separation-of-Dirac}Separation of Dirac equation in Elliptical
coordinates}
In this section we separate the Dirac equation in elliptical coordinates \cite{1990-Villalba}. Detailed calculation for separation of variables can be seen in Appendix~\ref{sov}. Dirac equation in (2+1)d is given as
\begin{equation}
[\gamma^{0}\partial_{t}+\gamma^{1}\partial_{x}+\gamma^{2}\partial_{y}]\psi=0\label{maindirac}\end{equation}
We define the Dirac matrix representation to be $\gamma^{0}=\tau_{2}$, $\gamma^{1}=i\tau_{1}$ and $\gamma^{2}=-i\tau_{3}$ , where $\tau_{i}$'s
are the Pauli matrices We define elliptical coordinates via
\begin{eqnarray}
x & = & f(\mu,\nu)=R\,\cosh\mu\,\cos\nu\\
y & = & g(\mu,\nu)=R\,\sinh\mu\,\sin\nu\end{eqnarray}
Here $\mu$ is the radial coordinate and $\nu$ is the angular coordinate. Here, we directly define the transformations that separate the Dirac equation in elliptical coordinates and leave the rigorous details to Appendix~\ref{sov}.
\begin{equation}
TS^{-1}[\gamma^{0}\partial_{t}+\frac{\widetilde{\gamma}^{1}(\mu,\nu)}{h}\partial_{\mu}+\frac{\widetilde{\gamma}^{2}(\mu,\nu)}{h}\partial_{\nu}]ST(ST)^{-1}\psi=0\label{sdirac1}\end{equation}
The product of transformation matrices S and T can be explicitly written out as
\begin{equation}
ST=\frac{1}{(\cosh\,\mu+\cos\,\nu)}\left(\begin{array}{cc}
\cos\,\frac{\nu}{2}\cosh\,\frac{\mu}{2} & -\sin\,\frac{\nu}{2}\sinh\,\frac{\mu}{2}\\
\sin\,\frac{\nu}{2}\sinh\,\frac{\mu}{2} & \cos\,\frac{\nu}{2}\cosh\,\frac{\mu}{2}\end{array}\right)\label{ST}\end{equation}
\begin{equation}
\psi=ST\: Y\end{equation}
Y is the transformed wavefunction spinor and is given by \begin{equation}
Y=\left(\begin{array}{c}
\alpha(\mu)B(\nu)\\
i\beta(\mu)A(\nu)\end{array}\right)\label{Y}\end{equation}
Each component of Y satisfies separable second order equation. We define a separation constant $\lambda^{2}$ such that
\begin{eqnarray}
(\partial_{\nu}^{2}-ikR\cos\nu+k^{2}R^{2}\sin^{2}\nu+\lambda^{2})A(\nu) & = & 0\label{eq:whea}\\
\ (\partial_{\nu}^{2}+ikR\cos\nu+k^{2}R^{2}\sin^{2}\nu+\lambda^{2})B(\nu) & = & 0\label{eq:wheb}\\
(\partial_{\mu}^{2}-ikR\cosh\mu+k^{2}R^{2}\sinh^{2}\mu-\lambda^{2})\alpha(\mu) & = & 0\label{eq:wheal}\\
(\partial_{\mu}^{2}+ikR\cosh\mu+k^{2}R^{2}\sinh^{2}\mu-\lambda^{2})\beta(\mu) & = & 0\label{eq:whebe}\end{eqnarray}
which is equivalent to 4 coupled first order equations that connect the upper and lower components of the wave function spinor
\begin{eqnarray}
(\partial_{\nu}-ikR\sin\nu)A(\nu) & = & i\lambda B(\nu)\label{eq:coupledang1}\\
(\partial_{\nu}+ikR\sin\nu)B(\nu) & = & i\lambda A(\nu)\label{eq:coupledang2}\\
(\partial_{\mu}-ikR\sinh\mu)\alpha(\mu) & = & \lambda\beta(\mu)\label{eq:couplerrad}\\
(\partial_{\mu}+ikR\sinh\mu)\beta(\mu) & = & \lambda\alpha(\mu)\label{eq:coupledrad1}\end{eqnarray}
The separated equations Eq.~(\ref{eq:whea}), (\ref{eq:wheb}), (\ref{eq:wheal}), and (\ref{eq:whebe}) are known as the radial and angular Whittaker Hill equations (WHE) and the upper and lower components of the spinor are connected via first order coupled equations.

\section{\label{sec:Solutions-to-Whittakker-Hill's}Solutions to the Whittakker Hill
equation (WHE)}
We transform the radial and angular WHE by using the following functional transform.
 \begin{equation}
A(\nu)=y_{1}e^{-ikR\,\cos\,\nu}\end{equation}
where $y_{1}$ satisfies the differential equation known as the Ince equation \cite{urwin,ince,arscott1967, arscottbook} .
\begin{equation}
y_{1}^{''}+2ikR\sin\,\nu\, y_{1}^{'}+\lambda^{2}y_{1}=0\label{eq:angince}\end{equation}
We do similar transformations for the radial WHE
\begin{equation}
\alpha(\mu)=x_{2}e^{ikR\,\cosh\,\mu}\end{equation}
to obtain
\begin{equation}
x_{2}^{''}+2ikR\sinh\,\mu\, x_{2}^{'}-\lambda^{2}x_{2}=0\label{eq:radince}\end{equation}
$B(\nu)$ and $\beta(\mu)$ are calculated from the first order coupled equations. Now we will try to solve all the Ince equations as an eigenvalue problem using a matrix method. We write a general form of the Ince equation from which we can deduce Ince equations (\ref{eq:angince}) and (\ref{eq:radince}) and try to find the recursions for the general equation \cite{urwin}.
\begin{equation}
\frac{d^{2}\psi}{d\theta^{2}}+2i\,\omega\,\sin\,\theta\frac{d\psi}{d\theta}+(\lambda^{2}+\omega(\rho+i)\cos\,\theta)\psi=0\label{gince}\end{equation}
\begin{center}
$\psi=y_{1}$, $\theta=\nu$, $\rho=-i$ and $\omega=kR$ will yield
Eq.~(\ref{eq:angince}).
\par\end{center}
\begin{center}
$\psi=x_{2}$, $\theta=i\mu$, $\rho=-i$ and $\omega=-kR$ will yield
Eq.~(\ref{eq:radince}).
\par\end{center}

\subsection{\label{sub:angular}Solutions to angular WHE (Matrix method)}

As often happens in the solution of differential equations with periodic coefficients, the solutions fall into four classes corresponding to the four types of fourier series. They may be even or odd functions of $\theta$ and may have 2$\pi$ as their period or antiperiod. The four possible solutions to Eq.~(\ref{gince}) are
\begin{eqnarray}
\psi^{1}_{m}(\theta) & = & \sum_{r=0}^{\infty}a_{mr}\,\cos r\theta,\qquad \ \psi^{2}_{m}(\theta)=\sum_{r=0}^{\infty}b_{mr}\,\sin r\theta\label{eq:2piper}\\
\psi^{3}_{m}(\theta) & = & \sum_{r=0}^{\infty}a_{mr}\,\cos(r+\frac{1}{2})\theta,\qquad\ \psi^{4}_{m}(\theta)=\sum_{r=0}^{\infty}b_{mr}\,\sin(r+\frac{1}{2})\theta\label{eq:4piper}\end{eqnarray}
We know that the full wave function has to have 2$\pi$ periodicity in $\nu$. In order to have this we have to choose the solutions of the Ince equations to be 2$\pi$ antiperiodic in $\nu$. This is so because the transformation matrix (ST) multiplying the wave function spinor is 2$\pi$ antiperiodic in $\nu$ (see Appendix~\ref{sov} for details of the transformation matrix), hence the product of antiperiodic spinor and antiperiodic transformation matrix will yield 2$\pi$ periodicity in $\nu$. Hence the possible solutions are Eq.~(\ref{eq:4piper}). Now we substitute these solutions in Eq.~(\ref{gince}) and get recursion relations for the coefficients. Reducing WHE to Ince equations has the advantage that the Ince equation has only three term recursion relations as opposed to the five term recursions for WHE.
The eigenvalue equation for the even solution is
\begin{eqnarray}
a_{0}(\lambda e^{2}-\frac{1}{4}+\frac{\omega\rho}{2})+a_{1}\frac{\omega}{2}(\rho-2i) & = & 0\,\,\,\,\,\,(r=0)\label{eq:evaleven1}\\
\frac{\omega}{2}(\rho+2r\, i)a_{r-1}+(\lambda e^{2}-(r+\frac{1}{2})^{2})a_{r}+\frac{\omega}{2}(\rho-2(r+1)i)a_{r+1} & = & 0\,\,\,\,\ \ \ (r\geq1)\label{eq:evaleven2}\end{eqnarray}
The eigenvalue equation for the odd solution is
\begin{eqnarray}
b_{0}(\lambda o^{2}-\frac{1}{4}-\frac{\omega\rho}{2})+b_{1}\frac{\omega}{2}(\rho-2i) & = & 0\,\,\,\,\,\,(r=0)\label{eq:evalodd1}\\
\frac{\omega}{2}(\rho+2r\, i)b_{r-1}+(\lambda o^{2}-(r+\frac{1}{2})^{2})b_{r}+\frac{\omega}{2}(\rho-2(r+1)i)b_{r+1} & = & 0\,\,\,\,\ \ \ (r\geq1)\label{eq:evalodd2}\end{eqnarray}
 We can see that the coefficients of $a_{0}$ and $b_{0}$ are different in the recursions for even and odd parity solutions. This implies that the cosine type series solution and sine series solution have different eigenvalues ($\lambda e_{m}^{2}$ and $\lambda o_{m}^{2}$ where m is the eigenvalue index) and they turn out to be complex conjugate to each other. The complex eigenvalue is a consequence of non-Hermiticity of the Whittaker Hill equation. These recursions can be written in a matrix form. The solutions can be expressed as an infinite trigonometric series and we have to truncate it at a point where the extra terms are not significant. We can obtain the eigenvalues and eigenvectors of the matrix which will provide the complete solutions to the Ince equations that are used in the solution to Whittaker Hill equations.
Summarizing the angular solutions we can write
\begin{eqnarray}
Ae_{m}(\nu) & = & e^{-ikR\,\cos\,\nu}\sum_{r=0}^{\infty}a_{mr}\cos(r+\frac{1}{2})\nu\\
Ao_{m}(\nu) & = & e^{-ikR\,\cos\,\nu}\sum_{r=0}^{\infty}b_{mr}\sin(r+\frac{1}{2})\nu\end{eqnarray}

$Ae_{m}(\nu)$ and $Ao_{m}(\nu)$ are solutions for the angular factor of the lower component of the separated spinor. Using the first order coupled equation (\ref{eq:coupledang1}) we can obtain the upper component angular eigenfunctions,
\begin{eqnarray}
Be_{m}(\nu) & = & \frac{1}{i\lambda e_{m}}(\partial_{\nu}-ikR\sin\nu)Ae_{m}(\nu)\\
Bo_{m}(\nu) & = & \frac{1}{i\lambda o_{m}}(\partial_{\nu}-ikR\sin\nu)Ao_{m}(\nu)
\end{eqnarray}
$Be_{m}(\nu)$ and $Bo_{m}(\nu)$ are of the opposite parity to $Ae_{m}(\nu)$ and $Ao_{m}(\nu)$ due to the above operation and are given as
\begin{eqnarray}
Be_{m}(\nu) & = & \frac{1}{i\lambda e_{m}}e^{-ikR\,\cos\,\nu}\sum_{r=0}^{\infty}a_{mr}(-r-\frac{1}{2})\sin(r+\frac{1}{2})\nu\\
Bo_{m}(\nu) & = & \frac{1}{i\lambda o_{m}}e^{-ikR\,\cos\,\nu}\sum_{r=0}^{\infty}b_{mr}(r+\frac{1}{2})\cos(r+\frac{1}{2})\nu
\end{eqnarray}
Note that for the sake of notation we always classify eigenfunctions according to the eigenvalues $\lambda e_{m}^{2}$ and $\lambda o_{m}^{2}$. Functions corresponding to $\lambda e_{m}^{2}$ get suffix "e" and corresponding to $\lambda o_{m}^{2}$ get the suffix "o".
\subsection{\label{sub:Radia}Solutions to radial WHE}

Since we have obtained the eigenvalues by solving the angular equations, the eigenvalues can be used as parameters in the radial differential equations.

The first method to evaluate the radial solutions is to replace $\nu\rightarrow i\mu$ and $(kR\rightarrow-kR)$ in the angular solutions (which is the same transformation that connects radial and angular WHE), the regular periodic (in $i\mu$) solutions
are denoted by Je and Jo.
\begin{eqnarray}
Je_{m}(\mu) & =e^{ikR\,\cosh\,\mu} & \sum_{r=0}^{\infty}c_{mr}\,\sinh(r+\frac{1}{2})\mu\\
Jo_{m}(\mu) & =e^{ikR\,\cosh\,\mu} & \sum_{r=0}^{\infty}d_{mr}\,\cosh(r+\frac{1}{2})\mu\end{eqnarray}
The lower component of the spinor can be obtained from coupled radial Eqs.~(\ref{eq:couplerrad}) and (\ref{eq:coupledrad1}). We denote the lower component radial solution with primes and keep this notation for all the lower component radial solutions. Note that prime does not imply derivative but is defined by the following operator acting on the upper component solutions.
\begin{eqnarray}
Je'_{m}(\mu) & = & \frac{1}{\lambda e_{m}}(\partial_{\mu}-ikR\sinh\mu)Je_{m}(\mu)\\
Jo'_{m}(\mu) & = & \frac{1}{\lambda o_{m}}(\partial_{\mu}-ikR\sinh\mu)Jo_{m}(\mu)
\end{eqnarray}
The second linearly independent solution that is non-periodic in ($i\mu$) is given as
\begin{eqnarray}
Ne_{m}(\mu) & = & C_{m}^{e}(kR)\mu Je_{m}(\mu)+C_{m}^{e}(kR)\sum_{r=0}^{\infty}f_{mr}\,\cosh(r+\frac{1}{2})\mu\\
No_{m}(\mu) & = & C_{m}^{o}(kR)\mu Jo_{m}(\mu)+C_{m}^{o}(kR)\sum_{r=0}^{\infty}g_{mr}\,\sinh(r+\frac{1}{2})\mu\end{eqnarray}

where the presence of factor of $\mu$ ensures the non-periodicity of the second solutions in $i\mu$. Note that it is of opposite parity to the regular Je and Jo. This approach is similar to the calculation of non-periodic second solutions of the modified Mathieu equation \cite{mclachlan}. $C_{m}^{e,o}(kR)$ are the normalization constants.

The second method to calculate the radial solutions is the power series method. In this method we simply do a power series analysis in $\mu$ for the second order radial Ince equation (\ref{eq:radince}) (note that such solutions do not capture the complex periodicity of the hyperbolic functions). We then immediately get the two independent solutions of even and odd parity with the predetermined eigenvalues from the angular solutions acting as a parameter characterizing the different radial solutions.
\begin{eqnarray}
\ \ Je_{m}(\mu) & = & e^{ikR\,\cosh\,\mu}\sum_{r=0}^{\infty}c_{r}\,\mu^{2r+1}\ ,\ \ Ne_{m}(\mu)=e^{ikR\,\cosh\,\mu}\sum_{r=0}^{\infty}f_{r}\,\mu^{2r}\\
Jo_{m}(\mu) & = & e^{ikR\,\cosh\,\mu}\sum_{r=0}^{\infty}d_{r}\,\mu^{2r},\ \ \ No_{m}(\mu)=e^{ikR\,\cosh\,\mu}\sum_{r=0}^{\infty}g_{r}\,\mu^{2r+1}\end{eqnarray}
We are interested in studying the scattering problem which requires the radial solutions to have a well defined asymptotic form. We can evaluate the radial solutions as power series in $\mu$ and in series of $\sinh\mu$ and $\cosh\mu$ as described in the above mentioned methods. But these forms of solution diverge at large $\mu$ and therefore do not yield proper asymptotic forms. Fortunately radial solutions to Whittaker Hill equations can be written as series of confluent hypergeometric functions \cite{figuerdo,figuerdo2,figuerdo3, Ronveaux} which converge for all $\mu$. We follow the procedure described in Ref.~(\onlinecite{figuerdo}). We start with radial Ince equation (\ref{eq:radince})
\begin{equation}
\alpha^{''}(\mu)+2ikR\sinh\,\mu\,\alpha^{'}(\mu)-\lambda^{2}\alpha(\mu)=0\end{equation}
and make the transformation $z=\cosh^{2}\frac{\mu}{2}$. The resulting equation takes the form,
\begin{equation}
z(z-1)\alpha^{''}(z)+(4ikR\ z^{2}-4ikR\ z+z-\frac{1}{2})\alpha^{'}(z)-\lambda^{2}\alpha(z)=0\end{equation}
To extract the even and odd parity of solutions, we make the following functional transformations to the above equation.
\begin{equation}
\alpha(z)=\sqrt{z}\ \alpha e(z)\ \mbox{(for\ even\ parity)},\ \alpha(z)=\sqrt{z-1}\ \alpha o(z)\ \mbox{(for\ odd\ parity)}\ \end{equation}
Making these transformations, we get the following differential equations for $\alpha e(z)$ and $\alpha o(z)$.
\begin{eqnarray}
z(z-1)\alpha e^{''}(z)+\frac{1}{2}(8ikR\ z(z-1)\ +4z-3)\alpha e^{'}(z)+(2ikR\ (z-1)+\frac{1}{4}-\lambda e^{2})\alpha e(z) & = & 0\label{eq:evenhyp}\\
z(z-1)\alpha o^{''}(z)+\frac{1}{2}(8ikR\ z(z-1)\ +4z-1)\alpha o^{'}(z)+(2ikR\ (z)+\frac{1}{4}-\lambda o^{2})\alpha o(z) & = & 0\label{eq:oddhyp}\end{eqnarray}
We have classified the eigenvalues as $\lambda e^{2}$ and $\lambda o^{2}$ for the even parity solutions and the odd parity solutions respectively (we know the eigenvalues from the angular eigenvalue Eqs.~(\ref{eq:evaleven1},\ref{eq:evaleven2}) and Eqs.~(\ref{eq:evalodd1},\ref{eq:evalodd2}).
Solutions to the above equations can be expressed in terms of confluent hypergeometric functions
\begin{eqnarray}
\alpha e_{m}(z) & = & \sum_{n=0}^{\infty}c_{m\ n}^{e}\, M(n+1/2,n+2,-4ikR\ z)\\
\alpha o_{m}(\mu) & = & \sum_{n=0}^{\infty}c_{m\ n}^{o}\, M(n+1/2,n+2,-4ikR\ z)\end{eqnarray}
where the M are the regular hypergeometric functions satisfying the Kummer differential equation\cite{grad}
\begin{equation}
zM''(z)+(b-z)M'(z)-aM(z)=0\end{equation}
The three term recursion relations for the coefficients $c_{m}^{e}$ and $c_{m}^{o}$ are
\begin{eqnarray}
c^{e}_{m\ n-1}4ikR(n-\frac{1}{2})^{2}+c^{e}_{m\ n}(n(n+1)+4ikR\ n+\frac{1}{4}-\lambda e_{m}^{2})+(n+1)c^{e}_{m\ n+1} & = & 0\\
c^{o}_{m\ n-1}4ikR(n-\frac{1}{2})(n+\frac{1}{2})+c^{o}_{m\ n}(n(n+1)+4ikR\ (n+\frac{1}{2})+\frac{1}{4}-\lambda o_{m}^{2})+(n+1)c^{o}_{m\ n+1} & = & 0\end{eqnarray}
And the full solution to the radial WHE (from the solution to the Ince equation) can be written as
\begin{eqnarray}
Jo_{m}(\mu) & = & e^{ikR\,\cosh\,\mu}\sqrt{\cosh^{2}\frac{\mu}{2}}\sum_{n=0}^{\infty}c^{e}_{m\ n}\, M(n+\frac{1}{2},n+2,-4ikR\cosh^{2}\frac{\mu}{2})\label{eq:jo}\\
Je_{m}(\mu) & = & e^{ikR\,\cosh\,\mu}\sqrt{\cosh^{2}\frac{\mu}{2}-1}\sum_{n=0}^{\infty}c^{o}_{m\ n}\, M(n+\frac{1}{2},n+2,-4ikR\cosh^{2}\frac{\mu}{2})\label{eq:je}\end{eqnarray}
The second linearly independent solution can be obtained from the first solution using the following method.
\begin{eqnarray}
Fey_{m}(\mu) & = & Je_{m}(\mu)\ \intop_{\mu_{0}}^{\mu}\frac{1}{Je_{m}(\mu')^{2}}d\mu'\label{eq:fey}\\
Gey_{m}(\mu) & = & Jo_{m}(\mu)\ \intop_{\mu_{0}}^{\mu}\frac{1}{Jo_{m}(\mu')^{2}}d\mu'\label{eq:gey}\end{eqnarray}
Fey and Gey are the second linearly independent solutions corresponding to Je and Jo. The lower component of the spinor can be evaluated by using the coupled equations (\ref{eq:couplerrad}).
\begin{eqnarray}
Fey'_{m}(\mu) & = & \frac{1}{\lambda e_{m}}(\partial_{\mu}-ikR\sinh\mu)Fey_{m}(\mu)\\
Gey'_{m}(\mu) & = & \frac{1}{\lambda o_{m}}(\partial_{\mu}-ikR\sinh\mu)Gey_{m}(\mu)\\\end{eqnarray}
Solutions to the radial WHE for both upper component ($\alpha e_{m}(\mu)$, $\alpha o_{m}(\mu)$) and lower component ($\beta e_{m}(\mu)$, $\beta o_{m}(\mu)$) can be summarized in the combination of two linearly independent solutions as,
\begin{eqnarray}
\alpha e_{m}(\mu) & = & A_{m}^{e}Je_{m}(\mu)+B_{m}^{e}Fey_{m}(\mu)\\
\beta e_{m}(\mu) & = & A_{m}^{e}Je'_{m}(\mu)+B_{m}^{e}Fey'_{m}(\mu))\\
\alpha o_{m}(\mu) & = & A_{m}^{o}Jo_{m}(\mu)+B_{m}^{o}Gey_{m}(\mu)\\
\beta o_{m}(\mu) & = & A_{m}^{o}Jo'_{m}(\mu)+B_{m}^{o}Gey'_{m}(\mu))\end{eqnarray}

$A_{m}^{e},\ B_{m}^{e}$ and $A_{m}^{o},\ B_{m}^{o}$ are the undetermined coefficients and for notation sake we classify undetermined coefficients according to eigenvalues corresponding to $\lambda e_{m}^{2}$ or $\lambda o_{m}^{2}$. We identify undetermined coefficients with superscript "e" and "o" corresponding to the eigenvalues. A normalized choice for the undetermined coefficients would be
\begin{eqnarray}
A_{m}^{e} & = & \cos\delta_{m}^{e} \qquad A_{m}^{o}=\cos\delta_{m}^{o}\\
B_{m}^{e} & = & \sin\delta_{m}^{e} \qquad B_{m}^{o}=\sin\delta_{m}^{o}
\end{eqnarray}
Such a choice is helpful in formulating the scattering cross section in terms of phase shifts in the scattering amplitude with $\delta_{m}^{e}$ an $\delta_{m}^{o}$ being the phase shifts.

Armed with all the solutions to the individual components of the separated Dirac spinor, we can now write the full solution to the free Dirac equation as a superposition of all the eigenstates of the separated equations
\begin{equation}
\psi(\mu,\nu)=(ST)\sum_{m}\left(\begin{array}{c}
\alpha e_{m}(\mu)Be_{m}(\nu)+\alpha o_{m}(\mu)Bo_{m}(\nu)\\
i(\beta e_{m}(\mu)Ae_{m}(\nu)+\beta o_{m}(\mu)Ao_{m}(\nu))\end{array}\right)\label{eq:fullsolution}\end{equation}

\section{\label{sec:Expansion-of-plane}Expansion of incoming plane wave spinor
in terms of Whittaker Hill eigenfunctions}

The form of the incoming plane wave (see Appendix~\ref{sec:BdG} Eq.~(\ref{eq:Phii})) is given as $e^{i\,\vec{k}\cdot\vec{r}}\left(\begin{array}{c} \cos\frac{\theta}{2}\\ \sin\frac{\theta}{2}\end{array}\right)$ ($\theta$ is the angle of incidence of the quasiparticle current). One of the requirements to construct the scattering cross section is to expand the incident plane wave in terms of eigenfunctions of the free Dirac equation which satisfy the following continuity condition (\ref{bnd}) at $\mu=0$.
\begin{equation} \psi(\mu,-\nu)\mid_{\mu=0}=\psi(\mu,\nu)\mid_{\mu=0}\\\label{bnd}\end{equation}
 To write the plane wave expansion we take the following path. We write the free solution of the Dirac equation in elliptical coordinates as a linear combination of the eigenstates. One specific superposition of these eigenstates represents the plane wave spinor. Our aim in this section is to obtain these linear combination coefficients which represent the plane wave spinor. We can see that applying condition (\ref{bnd}) to $\psi(\mu,\nu)$ is same as applying it to $Y(\mu,\nu)$ (see Eq.~(\ref{Y})). This is because the transformation matrix which connects $\psi$ to Y cancels on both sides of the Eq.~(\ref{bnd}). \begin{equation} Y(\mu,-\nu)\mid_{\mu=0}=Y(\mu,\nu)\mid_{\mu=0}\\\label{bnd2}\end{equation}
Applying the above condition, we can find the following constraint on the two undetermined constants (per eigenstate) appearing in the radial solutions due to the overall parity of the eigenfunctions.
\begin{equation}
B_{m}^{e}=B_{m}^{o}=0\end{equation}
Thus we see that the radial functions with constraint at $\mu=0$ do not depend on Fey, Gey . Hence, the plane wave term only has Je and Jo terms which is analogous to the plane wave expansion in terms of Bessel functions in polar coordinates (see Ref.~\onlinecite{grad}) which only contains regular J Bessel functions. Applying the appropriate boundary conditions on Y, we find the plane wave solution to be:
\begin{equation}
\psi=(ST)\sum_{m}\left(\begin{array}{c}
(A_{m}^{e}Je_{m}Be_{m}(\nu)+A_{m}^{o}Jo_{m}Bo_{m}(\nu))\\
(i\ A_{m}^{e}Je'_{m}Ae_{m}(\nu)+i\ A_{m}^{o}Jo'_{m}Ao_{m}(\nu))\end{array}\right)\end{equation}
The above solution with arbitrary constants $A_{m}^{e}$ and $A_{m}^{o}$ is an arbitrary superposition of eigenstates. We need to calculate the linear combination coefficients (as a function of $\theta$) for which the expansion represents a plane wave spinor (\ref{eq:Phii}). We write down the following expansion for the plane wave incident at an angle $\theta$ with respect to the x-axis. This step is important since we would like to control the angle of incidence of the incoming quasiparticle current.
\begin{equation}
e^{i\,\vec{k}\cdot\vec{r}}\left(\begin{array}{c}
\cos\frac{\theta}{2}\\
\sin\frac{\theta}{2}\end{array}\right)=(ST)\left\{ \sum_{m}n_{m}^{e}Be_{m}(\theta)\left(\begin{array}{c}
Je_{m}Be_{m}\\
i\, Je_{m}^{'}Ae_{m}\end{array}\right)+\sum_{m}n_{m}^{o}Bo_{m}(\theta)\left(\begin{array}{c}
Jo_{m}Bo_{m}\\
i\, Jo_{m}^{'}Ao_{m}\end{array}\right)\right\} \label{eq:planeexp}\end{equation}
For details of the calculation of the plane wave expansion coefficients $n_{m}^{e}$ and $n_{m}^{o}$ refer to Appendix~\ref{sec:Appendix--Plane-wave}.

\section{\label{sec-construction} scattering amplitude and phase shifts}

To study the scattering problem one requires well defined asymptotic forms of the full wavefunction spinor. The asymptotic form of the wavefunction determines the scattering amplitude in terms of the phase shifts of each partial wave with respect to the partial waves in the incident quasiparticle current.
Thus, we consider the asymptotic limit $\mu\rightarrow\infty$ for the full wave function spinor $\psi=(ST\ )Y$.
\begin{equation}
\psi(\mu,\nu)=\frac{1}{(\cosh\,\mu+\cos\,\nu)}\left(\begin{array}{cc}
\cos\,\frac{\nu}{2}\cosh\,\frac{\mu}{2} & -\sin\,\frac{\nu}{2}\sinh\,\frac{\mu}{2}\\
\sin\,\frac{\nu}{2}\sinh\,\frac{\mu}{2} & \cos\,\frac{\nu}{2}\cosh\,\frac{\mu}{2}\end{array}\right)\sum_{m}\left(\begin{array}{c}
\alpha e_{m}(\mu)Be_{m}(\nu)+\alpha o_{m}(\mu)Bo_{m}(\nu)\\
i(\beta e_{m}(\mu)Ae_{m}(\nu)+\beta o_{m}(\mu)Ao_{m}(\nu))\end{array}\right)\label{eq:fullsolutionas}\end{equation}
To proceed further we need to use asymptotic forms for the radial functions. As $\mu\rightarrow\infty$, the asymptotic form (see Appendix~ \ref{sec:Appendix--Asymptotic-form}) for the radial functions is given as
\begin{eqnarray}
\alpha e_{m}(\mu) & \sim & \sqrt{\frac{1}{\pi kR}}e^{-i\frac{\pi}{4}}(\cos\delta_{m}^{e}+i\sin\delta_{m}^{e})e^{i\frac{kR}{2}\, e^{\mu}}\,\label{asymtote1}\\
\beta e_{m}(\mu) & \sim & \sqrt{\frac{1}{\pi kR}}e^{-i\frac{\pi}{4}}(\cos\delta_{m}^{e}-i\sin\delta_{m}^{e})e^{-i\frac{kR}{2}\, e^{\mu}}\,\\
\alpha o_{m}(\mu) & \sim & \sqrt{\frac{1}{\pi kR}}e^{-i\frac{\pi}{4}}(\cos\delta_{m}^{o}+i\sin\delta_{m}^{e})e^{i\frac{kR}{2}\, e^{\mu}}\,\\
\beta o_{m}(\mu) & \sim & \sqrt{\frac{1}{\pi kR}}e^{-i\frac{\pi}{4}}(\cos\delta_{m}^{o}-i\sin\delta_{m}^{e})e^{-i\frac{kR}{2}\, e^{\mu}}\,\end{eqnarray}
For compact notation, we can write $e^{\mu}=\frac{2r}{R}$, where R is half the distance between the foci of the elliptical coordinates and r is the polar radial coordinate in the limit where the angular elliptical coordinate $\nu$ approaches the polar angle $\phi$. In large $\mu$ limit elliptical coordinates reduce to polar coordinates.
\begin{eqnarray}
\alpha e_{m}(r) & \sim & \sqrt{\frac{1}{\pi kR}}e^{-i\frac{\pi}{4}}e^{i(kr+\delta_{m}^{e})}\,\label{asymtote2}\\
\beta e_{m}(r) & \sim & \sqrt{\frac{1}{\pi kR}}e^{-i\frac{\pi}{4}}e^{-i(kr+\delta_{m}^{e})}\,\\
\alpha o_{m}(r) & \sim & \sqrt{\frac{1}{\pi kR}}e^{-i\frac{\pi}{4}}e^{i(kr+\delta_{m}^{o})}\,\\
\beta o_{m}(r) & \sim & \sqrt{\frac{1}{\pi kR}}e^{-i\frac{\pi}{4}}e^{-i(kr+\delta_{m}^{o})}\,\end{eqnarray}
The asymptotic form of the ST transformation matrix can also be evaluated. As $\mu\rightarrow\infty$, $\cosh\mu+\cos\nu\sim\cosh\mu$ and $\cosh\mu\sim\sinh\mu\sim\frac{e^{\mu}}{2}=\frac{r}{R}$. Applying these limits in the ST matrix of Eq.~(\ref{ST}) gives
\begin{eqnarray}
ST & \sim & \frac{\sqrt{R}}{\sqrt{r}}\left(\begin{array}{cc}
\cos\frac{\nu}{2} & -\sin\frac{\nu}{2}\\
\sin\frac{\nu}{2} & \cos\frac{\nu}{2}\end{array}\right)\label{eq:asymst}\end{eqnarray}
We can then substitute the asymptotic forms of the radial solutions back into the full solution spinor of Eq.~(\ref{eq:fullsolution}) to obtain
\begin{equation}
\psi=(ST)\sqrt{\frac{1}{\pi kR}}e^{-i\frac{\pi}{4}}\sum_{m}\left\{ e^{ikr}(e^{i\delta_{m}^{e}}Be_{m}(\nu)+e^{i\delta_{m}^{o}}Bo_{m}(\nu))\left(\begin{array}{c}
1\\
0\end{array}\right)+e^{-ikr}i(e^{-i\delta_{m}^{e}}Ae_{m}(\nu)+e^{-i\delta_{m}^{o}}Ao_{m}(\nu))\left(\begin{array}{c}
0\\
1\end{array}\right)\right\} \label{eq:full soln phaseshift}\end{equation}
We would like to write the full wave function in the two suggestive parts requisite to set up the scattering problem
\begin{equation}
\psi=\psi^{plane\, wave}+\psi^{scattered}\label{eq:suggestive}\end{equation}
The plane wave expanded in terms of the separated eigenstates Eq.~(\ref{eq:planeexp}) in the large $\mu$ limit takes the form
\begin{eqnarray}
\psi^{plane\, wave} & = & (ST)\sqrt{\frac{1}{\pi kR}}e^{-i\frac{\pi}{4}}\sum_{m}e^{ikr}(n_{m}^{e}Be_{m}(\theta)Be_{m}(\nu)+n_{m}^{o}Bo_{m}(\theta)Bo_{m}(\nu))\left(\begin{array}{c}
1\\
0\end{array}\right)\label{eq:plane}\\
 & + & (ST)\sqrt{\frac{1}{\pi kR}}e^{-i\frac{\pi}{4}}\sum_{m}e^{-ikr}i(n_{m}^{e}Be_{m}(\theta)Ae_{m}(\nu)+n_{m}^{o}Bo_{m}(\theta)Ao_{m}(\nu))\left(\begin{array}{c}
0\\
1\end{array}\right)\end{eqnarray}
Now we construct the outgoing radial wave with appropriate asymptotic form.
\begin{equation}
\psi^{scattered}=(ST)\sum_{m}(d_{m}^{e}n_{m}^{e}Be_{m}(\theta)Be_{m}(\nu)He_{m}(\mu)+d_{m}^{o}n_{m}^{o}Bo_{m}(\theta)Bo_{m}(\nu)Ho_{m}(\mu))\left(\begin{array}{c}
1\\
0\end{array}\right)\label{eq:scattered}\end{equation}
$He_{m}(\mu)$ and $Ho_{m}(\mu)$ are the linear combinations of two linearly independent solutions to the radial WHE ($Je_{m}(\mu)$, $Fey_{m}(\mu)$) and ($Jo_{m}(\mu)$, $Gey_{m}(\mu)$) which have the behavior of an outgoing radial wave ($He_{m}(r) \sim Ho_{m}(r) \sim e^{ikr}$) in the asymptotic limit. They play the role of Hankel functions\cite{grad} (linear combination of Bessel J and Bessel Y) which appear in the study of scattering problems in polar coordinates. $d_{m}^{e}$ and $d_{m}^{o}$ are the undetermined coefficients. The asymptotic form of the scattered wave is given as
\begin{equation}
\psi^{scattered}(\mu\rightarrow\infty)=(ST)\sqrt{\frac{1}{\pi kR}}e^{-i\frac{\pi}{4}}e^{ikr}\sum_{m}(d_{m}^{e}n_{m}^{e}Be_{m}(\theta)Be_{m}(\nu)+d_{m}^{o}n_{m}^{o}Bo_{m}(\theta)Bo_{m}(\nu))\left(\begin{array}{c}
1\\
0\end{array}\right)\end{equation}
Now we compare our full wave function spinor, Eq.~(\ref{eq:fullsolution}) and with the wave functions written in a suggestive form in the asymptotic limit. $e^{ikr}$ and $e^{-ikr}$ multiplied by the angular functions in $\nu$ are independent functions. Hence we can equate their coefficients in Eq.~(\ref{eq:suggestive}) and Eq.~(\ref{eq:full soln phaseshift}). We get four equations for the undetermined coefficients corresponding to the four independent angular
functions.
\begin{eqnarray}
e^{i\delta_{m}^{e}} & = & (1+d_{m}^{e})n_{m}^{e}Be_{m}(\theta)\label{eq:coeff1}\\
e^{i\delta_{m}^{o}} & = & (1+d_{m}^{o})n_{m}^{o}Bo_{m}(\theta)\label{eq:coeff2}\\
e^{-i\delta_{m}^{e}} & = & (n_{m}^{e})Be_{m}(\theta)\label{eq:coeff3}\\
e^{-i\delta_{m}^{o}} & = & (n_{m}^{o})Bo_{m}(\theta)\label{eq:coeff4}\end{eqnarray}
Solving the above four equations we can write $d_{m}^{e}$ and $d_{m}^{o}$ (which are the undetermined coefficients of the scattered wave) in terms of the phase shifts $\delta_{m}^{e}$and $\delta_{m}^{o}$.
\begin{equation}
d_{m}^{e}=(e^{2i\delta_{m}^{e}}-1)\ ,\ d_{m}^{o}=(e^{2i\delta_{m}^{o}}-1)\ \label{eq:phaseshift}\end{equation}
To write the full form of the scattering amplitude we multiply the scattered wave by the ST transformation matrix in its asymptotic form.
\begin{equation}
\psi^{scattered}=\sqrt{\frac{1}{\pi k}}e^{-i\frac{\pi}{4}}\sum_{m}(d_{m}^{e}n_{m}^{e}Be_{m}(\theta)Be_{m}(\nu)+d_{m}^{o}n_{m}^{o}Bo_{m}(\theta)Bo_{m}(\nu))\frac{e^{ikr}}{\sqrt{r}}\left(\begin{array}{c}
\cos\frac{\nu}{2}\\
\sin\frac{\nu}{2}\end{array}\right)\end{equation}

$\left(\begin{array}{c}
\cos\frac{\nu}{2}\\
\sin\frac{\nu}{2}\end{array}\right)$ represents the quasiparticle current going in the radial direction
(see Eq.~(\ref{eq:jscat})). The scattering amplitude can be extracted from the asymptotic form of the scattered wave and is given as \begin{equation}
f(\theta,\nu)=\sqrt{\frac{1}{\pi k}}e^{-i\frac{\pi}{4}}(\sum_{m}(e^{2i\delta_{m}^{e}}-1)n_{m}^{e}Be_{m}(\theta)Be_{m}(\nu)+(e^{2i\delta_{m}^{o}}-1)\ n_{m}^{o}Bo_{m}(\theta)Bo_{m}(\nu))\label{eq:scattampdbl}\end{equation}

Hence, we were successful in constructing the scattering amplitude (analogous to the general form of scattering amplitude (Eq.~(\ref{eq:Phis})) in elliptical coordinates. The only thing that remains is to calculate the phase shifts $\delta_{m}^{e}$ and $\delta_{m}^{o}$, and for that we need to impose the conditions for the branch cut on the full wave function spinor, Eq.~(\ref{eq:fullsolution}).

\section{Scattering cross section without branch cut (Berry phase parameter B=0)}
\label{sanity}

Before going on to the case with the branch cut, we make a quick check on our scattering amplitude for the case of no branch cut or B=0 in Eq.~(\ref{bound}). We expect this trivial case to yield no scattering of quasiparticles.
For the case of no branch cut between the foci of the ellipse ($\mu=0$), we impose the following condition on the wave function spinor
\begin{equation} \psi(\mu,-\nu)\mid_{\mu=0}=\psi(\mu,\nu)\mid_{\mu=0} \label{trivialcond}\end{equation}
At $\mu=0$, we have the following behavior for $Je_{m}(\mu)$, $Je'_{m}(\mu)$, $Jo_{m}(\mu)$ and $Jo'_{m}(\mu)$
\begin{eqnarray}
Jo_{m}(0)\neq 0,\ Je_{m}(0)& = & 0\\
Je'_{m}(0)\neq 0,\ Jo'_{m}(0)& = & 0
\end{eqnarray}
At $\mu=0$, the second independent solutions are all nonzero,
\begin{eqnarray}
Fey_{m}(0)\neq 0,\ Gey_{m}(0)\neq 0\\
Fey'_{m}(0)\neq0,\ Gey'_{m}(0)\neq 0
\end{eqnarray}
Applying condition (\ref{trivialcond}), all the terms containing even parity angular eigenfunctions cancel out and we can write the remaining terms as
\[
\left(\begin{array}{c}
(\cos\delta_{m}^{e}Je_{m}(0)+\sin\delta_{m}^{e}Fey_{m}(0))Be_{m}(\nu)\\
i\ (\cos\delta_{m}^{o}Jo'_{m}(0)+\sin\delta_{m}^{o}Gey'_{m}(0))Ao_{m}(\nu)\end{array}\right)=\left(\begin{array}{c}
0\\
0\end{array}\right)\]
Substituting for the values of the radial functions at $\mu=0$, we have
\[
\left(\begin{array}{c}
(\sin\delta_{m}^{e}Fey_{m}(0))Be_{m}(\nu))\\
i\ (\sin\delta_{m}^{o}Gey'_{m}(0))Ao_{m}(\nu)\end{array}\right)=\left(\begin{array}{c}
0\\
0\end{array}\right)\]
From the boundary condition at the origin we get the following constraints on the undetermined coefficients,
\begin{equation}
\sin\delta_{m}^{o}=0,\ \ \ \sin\delta_{m}^{e}=0\end{equation}
The above expression gives phase shifts as $\delta_{m}^{e}=0$ and $\delta_{m}^{o}=0$. Putting the obtained phase shifts back into Eq.~(\ref{eq:scattampdbl}) for the scattering amplitude, we obtain \begin{equation}
f(\theta,\nu)=0 \end{equation}
Hence, we recover our trivial result that without the branch cut (and without superflow) there is no scattering. Now we move to the interesting case of quasiparticle scattering with the branch cut.

\section{\label{sec:Scattering-cross-section}Scattering cross section due to a branch cut (B=1)}

For the case with a branch cut between the foci of the ellipse ($\mu=0$), we set the Berry phase parameter B=1 which sets the condition imposed on the wave function spinor Eq.~(\ref{eq:fullsolution}) to,
 \begin{equation}
\psi(\mu,-\nu)\mid_{\mu=0}=-\psi(\mu,\nu)\mid_{\mu=0}\end{equation}
Applying the above condition and using the values of the radial eigenfunctions at $\mu=0$, all the terms containing odd angular eigenfunctions cancel out and we can write the remaining terms as
\[
\left(\begin{array}{c}
(\cos\delta_{m}^{o}Jo_{m}(0)+\sin\delta_{m}^{o}Gey_{m}(0))Bo_{m}(\nu)\\
i\ (\cos\delta_{m}^{e}Je'_{m}(0)+\sin\delta_{m}^{e}Fey'_{m}(0))Ae_{m}(\nu)\end{array}\right)=\left(\begin{array}{c}
0\\
0\end{array}\right)\]
From the boundary condition at the origin we get the following constraints on the undetermined coefficients,
\begin{equation}
\cos\delta_{m}^{o}Jo_{m}(0)=-\sin\delta_{m}^{o}Gey_{m}(0),\ \ \ \cos\delta_{m}^{e}Je'_{m}(0)=-\sin\delta_{m}^{e}Fey'_{m}(0)\end{equation}
In other words
\begin{eqnarray}
\tan\delta_{m}^{e} & = & -\frac{Je'_{m}(0)}{Fey'_{m}(0)}\label{eq:bcond1}\\
\tan\delta_{m}^{o} & = & -\frac{Jo_{m}(0)}{Gey_{m}(0)}\label{eq:bccond2}\end{eqnarray}
Since $\tan\delta_{m}^{e}$ and $\tan\delta_{m}^{o}$ are complex conjugates of each other we have to account for the relative sign between $\delta_{m}^{e}$ and $\delta_{m}^{o}$ while calculating the inverse tangent in the above relation. Substituting for $\delta_{m}^{e}$ and $\delta_{m}^{o}$ in the definitions of the phase shifts, Eq.~(\ref{eq:phaseshift}) we obtain
 \begin{eqnarray}
(e^{2i\delta_{m}^{e}}-1) & = & \frac{-2Je'_{m}(0)\ }{Je'_{m}(0)+i\ Fey'_{m}(0)}\label{finalshifte}\\
(e^{2i\delta_{m}^{o}}-1) & = & \frac{-2Jo_{m}(0)\ }{Jo_{m}(0)-i\ Gey_{m}(0)}\label{finalshifto}\end{eqnarray}

Using Eq.~(\ref{finalshifte}) and Eq.~(\ref{finalshifto}) we can completely evaluate the scattering amplitude and the differential cross section for quasiparticle scattering due to branch cut without the superflow.
\begin{equation}
f(\theta,\nu)=\sqrt{\frac{4}{\pi k}}e^{-i\frac{\pi}{4}}\sum_{m} (\frac{Je'_{m}(0)\ }{Je'_{m}(0)+i\ Fey'_{m}(0)})n_{m}^{e}Be_{m}(\theta)Be_{m}(\nu)+(\frac{Jo_{m}(0)\ }{Jo_{m}(0)-iGey_{m}(0)})n_{m}^{o}Bo_{m}(\theta)Bo_{m}(\nu)\label{eq:scattampdblfinal1}\end{equation}
We can write the exact differential cross section for the quasiparticle scattering in terms of $\varphi=\nu-\theta$,
\begin{equation}
\frac{d\sigma}{d\varphi}=\frac{4}{\pi k}\left|\sum_{m}(\frac{Je'_{m}(0)\ }{Je'_{m}(0)+i\ Fey'_{m}(0)})n_{m}^{e}Be_{m}(\theta)Be_{m}(\varphi+\theta)+(\frac{Jo_{m}(0)\ }{Jo_{m}(0)-i\ Gey_{m}(0)})n_{m}^{o}Bo_{m}(\theta)Bo_{m}(\varphi+\theta)\right|^{2}\label{eq:scattampdblfinal2}\end{equation}

Recall that at the outset of this calculation, we shifted the origin of momentum space to the center of node 1. Thus, in the discussions that followed, we have been considering quasiparticles scattered from one state in the vicinity of node 1 to another state in the vicinity of node 1. The resulting cross section is therefore only the cross section for these node-1 quasiparticles. However, given a quasiparticle current in any particular direction, quasiparticles from all four nodes will contribute equally. Thus to obtain the physical cross section, we must average over the cross sections for quasiparticles at each of the four nodes. Our results for node 1 can be easily generalized to node $j=\{1,2,3,4\}$ by transforming coordinates to those appropriate to node $j$. In accordance with the $d$-wave structure of the gap, we can define a local coordinate system at each of the four nodes with a $\hat{\bf k}_{1}$ axis pointing along the direction of increasing $\epsilon_{k}$ and a $\hat{\bf k}_{2}$ axis pointing along the direction of increasing $\Delta_{k}$. Note that while nodes 1 and 3 define right-handed coordinate systems, nodes 2 and 4 define left-handed coordinate systems. We can therefore transform from node 1 to node $j$ simply by rotating our incident and scattered angles ($\theta$ and $\nu$) and then changing the sign of these angles to account for the handedness of the local coordinate system.

\begin{equation}
\begin{array}{llll}
\mbox{Node 1:} & \theta_{1}=\theta
& \nu_{1}=\nu
& \varphi_{1} = \nu_{1}-\theta_{1} = \varphi \\
\mbox{Node 2:} & \theta_{2}=-(\theta-\mbox{$\frac{\pi}{2}$})
& \nu_{2}=-(\nu-\mbox{$\frac{\pi}{2}$})
& \varphi_{2} = \nu_{2}-\theta_{2} = -\varphi \\
\mbox{Node 3:} & \theta_{3}=\theta+\pi
& \nu_{3}=\nu+\pi
& \varphi_{3} = \nu_{3}-\theta_{3} = \varphi \\
\mbox{Node 4:} & \theta_{4}=-(\theta+\mbox{$\frac{\pi}{2}$})
& \nu_{4}=-(\nu+\mbox{$\frac{\pi}{2}$})
& \varphi_{4} = \nu_{4}-\theta_{4} = -\varphi
\end{array}
\label{eq:angles}
\end{equation}
Thus, to obtain results for quasiparticles about node j, we need only input each $\theta_{j}$ and take the output
as a function of $(-1)^{j+1} \varphi$. Then the physical cross sections are
\begin{equation}
\frac{d\sigma}{d\varphi} = \frac{1}{4}\sum_{j=1}^{4}
\left( \frac{d\sigma}{d\varphi} \right)_{j}
\;\;\;\;\;\;
\sigma_{\parallel} = \frac{1}{4}\sum_{j=1}^{4} \sigma_{\parallel}^{j}
\label{eq:4nodeaverage}
\end{equation}

\section{Differential cross section results for Berry phase scattering}
\label{results}

\begin{figure}[H]
\noindent \begin{centering}
\includegraphics[scale=0.39]{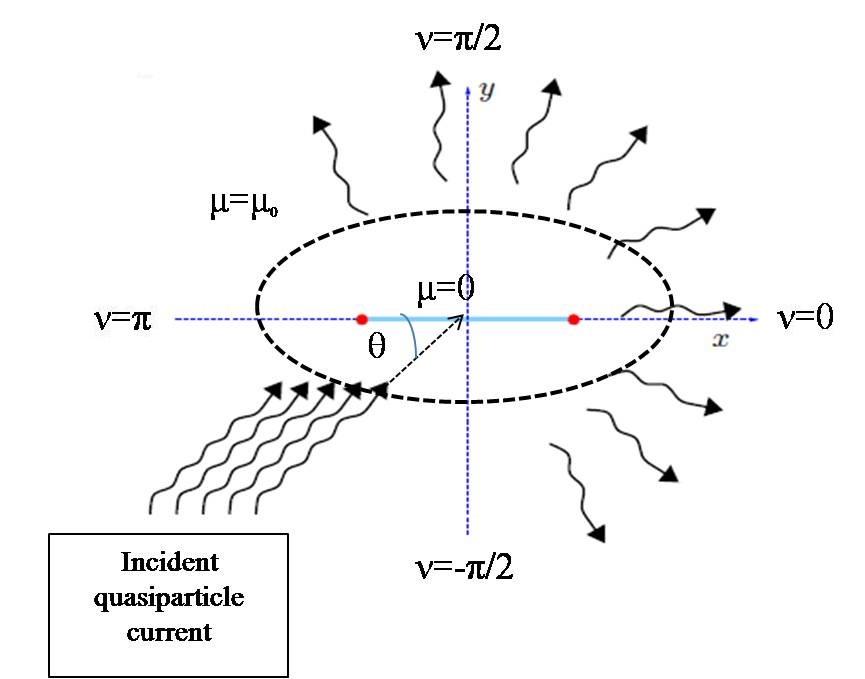}
\caption{The above picture depicts the scattering of quasiparticles due to the Berry phase effect. The Berry phase effect is denoted by the finite branch cut shown by the thick line joining the dots. The dots represent the vortex cores coinciding with the foci. Wiggly lines denote the incident quasiparticle current. $\theta$ is the incident angle of the quasiparticle current with respect to the x-axis. }
\label{fig:Schematics-of-scattering}
\par\end{centering}
\end{figure}

In this section we plot the differential cross section for quasiparticle scattering from branch cut (no superflow) for several cases. We define the distance between the two vortices by the dimensionless parameter kR. Incident angle for the current is described by $\theta$ as shown in Fig.~ \ref{fig:Schematics-of-scattering}. The thick dots on the foci of the ellipse depicts the vortex cores and the thick line joining the cores denotes the branch cut and is also the $\mu$=0 line in elliptical coordinates. The wiggly arrows represent the incident quasiparticle current. The angle of incidence is $\theta$ with respect to the x-axis. The plane wave spinor representing quasiparticle current is incident on the vortex cores and part of it acquires a Berry phase factor of (-1) between the vortex cores. The scattering contribution is entirely due to this effect.

\begin{figure}[H]
\noindent \begin{centering}
\includegraphics[scale=0.5]{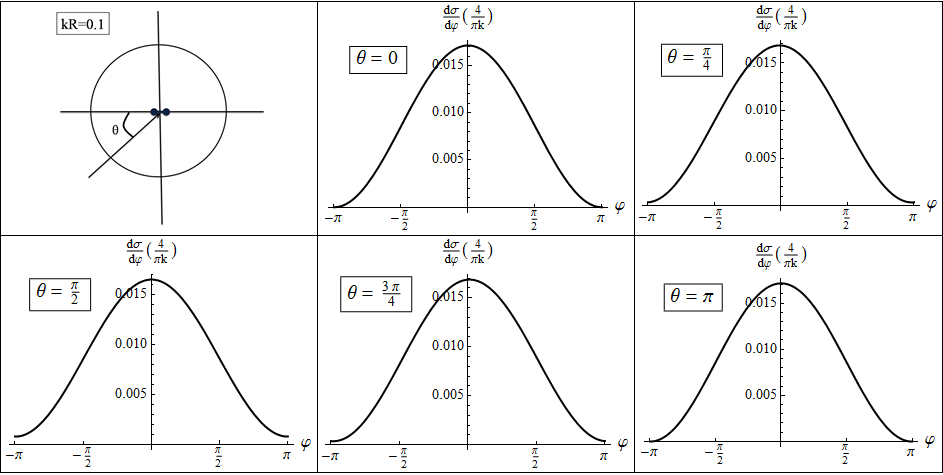}
\caption{\label{fig:smallkr}In the above figure vortex cores (foci of ellipse) are depicted by dots and the line joining them is the branch cut. Vortex cores are separated by dimensionless length kR=0.1. $\theta$ is the angle of incidence of the quasiparticle current. For small inter-vortex separation, the ellipse looks like a circle which indicates near-circular symmetry in the scatterer. We plot single node differential scattering cross section for quasiparticle current incident at different angles $\theta$. The plots of the scattering cross section emphasize the near-circular symmetry with respect to the incident angle $\theta$ due to small inter vortex separation.}
\label{fig:kr01}
\par\end{centering}
\end{figure}

\begin{figure}[H]
\noindent \begin{centering}
\includegraphics[scale=0.5]{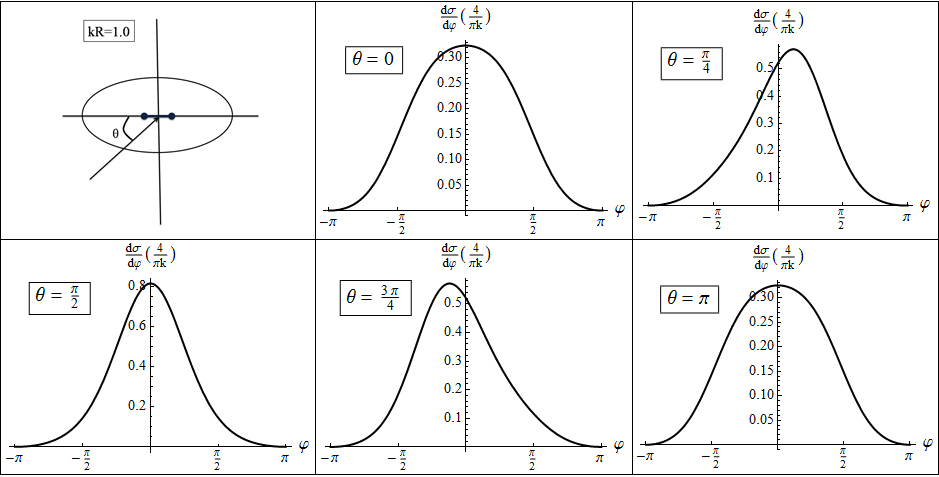}
\caption{\label{fig:kr1}In the above figure vortex cores are further apart with dimensionless length kR=1.0. With the increase in inter-vortex separation the scatterer becomes more elliptical and plots show expected elliptical symmetry in the single node differential scattering cross section. Also note the increased magnitude of scattering cross section which can be attributed to the increase in the length of branch cut. In other words, more quasiparticles hit the branch cut}
\label{fig:krone}
\par\end{centering}
\end{figure}

For very small inter vortex separation (kR=0.1) the ellipse looks like a circle (see Fig.~\ref{fig:smallkr}). In this limit we expect near-circular symmetry in the plots of differential cross section. We see that the quasiparticles see relatively small branch cut which results in the differential cross section being almost independent of the incident angle $\theta$ (see Fig.~\ref{fig:kr01}). Note that this is not a very good limit physically since we can no longer ignore the presence of other vortices in the sample. When we stretch the vortex cores apart (kR=1) the scatterer becomes more elliptical (see Fig.~\ref{fig:kr1}). This is reflected in the elliptical symmetry we see in the cross section plots as we rotate the incident angle of the quasiparticle current (see Fig.~\ref{fig:krone}). We see that we get the same plots for differential cross section if we rotate the incident angle by $\pi$, which reflects the symmetry of the scatterer (symmetric under $\pi$ rotation).

\begin{figure}[H]
\noindent \begin{centering}
\includegraphics[scale=0.5]{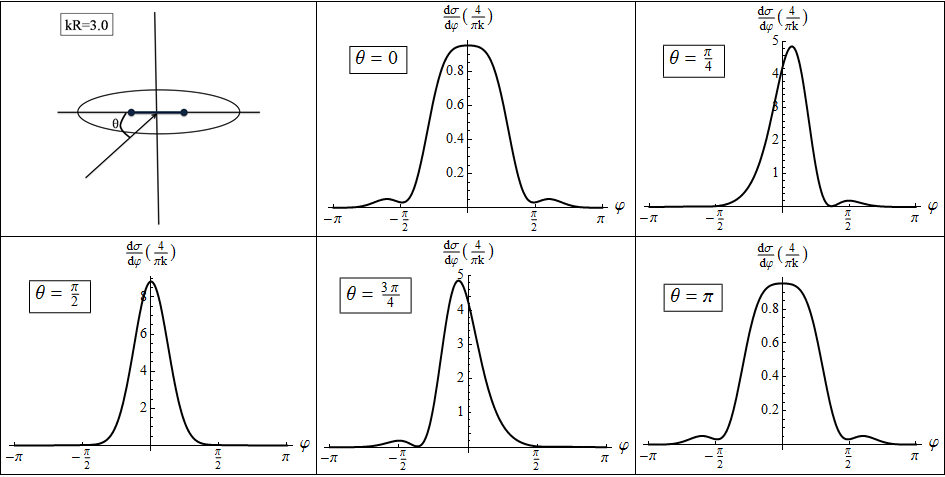}
\caption{In the above figure vortices are further apart with kR=3.0. The plots of the single node scattering cross section show elliptical symmetry. We also see increase in the magnitude of scattering cross section as compared to the case of kR=1.0}
\label{fig:krthreepointzero}
\par\end{centering}
\end{figure}
As we increase the inter-vortex separation further to kR=3.0, we obtain the case of a highly elliptical scatterer. For this case the magnitude of the scattering cross section increases as compared to the case of kR=1 (see Fig.~\ref{fig:krthreepointzero}). We observe the scattered current sweeping closer to the forward direction for higher kR. For kR=1 and kR=3, we see maximum scattering for the case of $\theta=\pi/2$. At this angle the quasiparticle current is normally incident on the branch cut and results in maximum exposure to the Berry phase effect. Mirror symmetry about $\theta=\pi/2$ is seen in the cross section plots. Thus our results for the scattering cross section are consistent with the geometry of the scatterer. We note that the scattering is reflectionless, or in other words, there is no backscattering of the quasiparticle current. Absence of backscattering due to the Berry phase has been previously reported in the literature for the case of carbon nanotubes (see Ref.~\onlinecite{ando}).

\begin{figure}[H]
\begin{centering}
\includegraphics[scale=0.5]{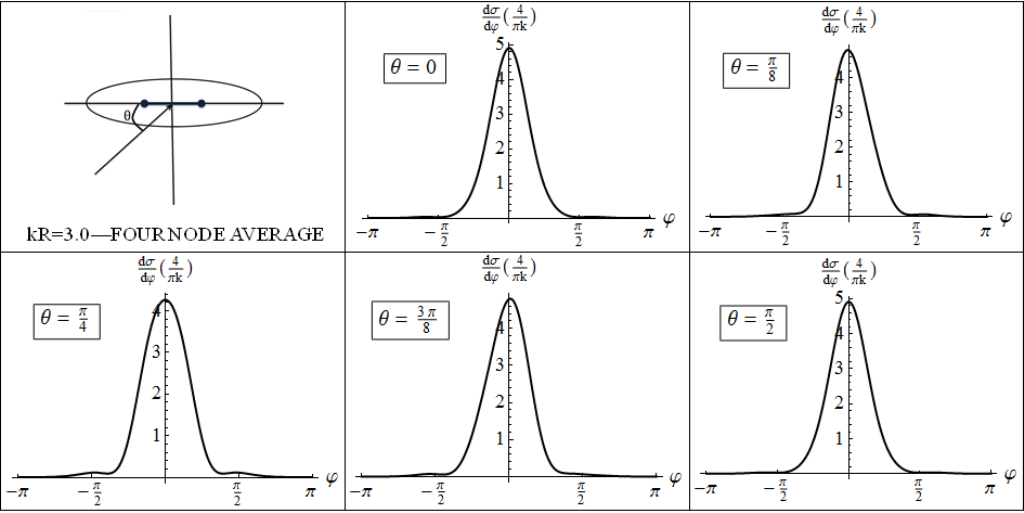}\\
\caption{ Four Node Average Differential Cross Section.  We plot differential cross section averaged over the contributions of quasiparticles from all four gap nodes, for quasiparticle current incident at various angles $\theta$ and for inter-vortex separation kR=3.0.  Results are $\pi/2$ periodic with respect to $\theta$. }
\label{fig:nodeavgbc}
\end{centering}
\vspace{0.5cm}
\begin{centering}
\includegraphics[scale=0.5]{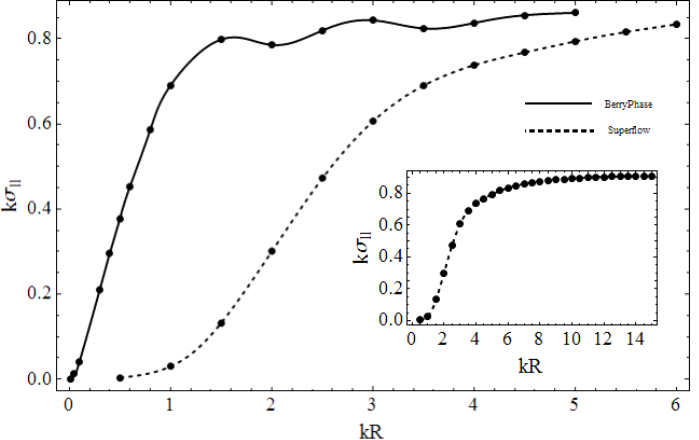}
\caption{$k\sigma_{\|}$ plotted versus increasing inter vortex separation, averaged over the incident angle $\theta$. The solid curve shows the transport cross section for the Berry Phase scattering case. Dashed curve shows the transport cross section for the superflow scattering. Inset shows $k\sigma_{\|}$ plot for the case of superflow scattering of quasiparticles plotted for very high kR values \cite{superflow1}.}
\label{fig:sigmadvp}
\end{centering}
\end{figure}

We must now average over the scattering contribution due to quasiparticles from all four nodes. The four node average has been performed as prescribed in Eq.~(\ref{eq:angles}) and Eq.~(\ref{eq:4nodeaverage}). After averaging over four nodes, we still see $\theta$ dependence in the differential cross section (see Fig.~\ref{fig:nodeavgbc}). The resulting cross section is $\pi/2$ periodic with respect to $\theta$. This is a consequence of the definition of $\theta$ at each node (see Eq.~(\ref{eq:angles})). We should keep in mind that in the setup we consider (cuprate sample), the pair of vortices are not always aligned along the x-axis as shown in Fig.~\ref{fig:Schematics-of-scattering}.

In Fig.~\ref{fig:sigmadvp}, we plot the total transport cross section, $k\sigma_{\|}$ as a function of inter-vortex separation kR.
\begin{equation}
k\sigma_{\parallel} = \int_{-\pi}^{\pi} \!d\varphi\, \frac{d\sigma}{d\varphi}
(1-\cos\varphi)
\label{eq:transCSdef}
\end{equation}
We notice that the transport cross section goes to 0 as kR$\rightarrow$0. This is expected since the Berry phase effect (branch cut) is negligible for very small values of kR. With the increase in kR, the longitudinal cross section increases rapidly and then saturates for kR>1. On the basis of the transport cross section plots obtained for the case of superflow and Berry phase scattering, one can make an intuitive comparison of these two effects. In the superflow paper \cite{superflow1}, we have neglected the Berry phase effect by applying periodic boundary conditions to the quasiparticle wave functions. This amounts to scattering of quasiparticles from vortices with superflow potential with the strength of two vortices. To calculate the transport cross section for the case of a branch cut between the two vortices, we have neglected the superflow contribution. Hence we have neatly isolated the scattering contributions due to these two effects, which gives us an opportunity to compare these two effects. Before such a comparison, we must treat the case of the Berry phase scattering on an equal footing with the superflow scattering. Due to the two-center nature of the regularized Berry phase effect, we are dealing with elliptical geometry in this case. Upon performing the four-node average, we see that this elliptical symmetry shown in the differential cross section plots (See Fig.~\ref{fig:krthreepointzero}) has been reduced to near circular symmetry even for the highly elliptical case of kR=3 (Fig.~\ref{fig:nodeavgbc}). Also, since there is no preferred orientation of the branch cut, one can average over the alignment of the branch cut with respect to the x-axis. The final DCS averaged over this alignment will have no elliptical symmetry or skew scattering. Based on the above arguments, we may directly compare the transport cross section due to superflow potential of two vortices of radii kR on top of each other to the Berry phase scattering due to two vortices separated by distance kR (averaged over incident angle $\theta$). In both cases, kR parameterizes the dimensionless energy of incident quasiparticles. kR also determines the size of the vortex for the case of superflow scattering and the length of the branch cut for the case of  Berry phase scattering. For the superflow case, we see a steep increase in the transport cross section (after averaging over 4 nodes) followed by saturation for kR > 5.   For the case of Berry phase scattering, the increase in transport cross section is steeper than for the superflow case. For the Berry phase scattering the magnitude of $k\sigma_{\|}$ saturates for kR > 1. This shows that the Berry phase is the more important effect of the two for kR < 5 (high field low temperature regime). Magnitudes of transport cross sections for higher kR (weak field high temperature regime) are similar for both the superflow and Berry phase processes. On the basis of the plots of transport cross section for both cases, one can conclude that the transport cross section due to the branch cut dominates for kR < 5, and is of similar order to the superflow contribution for kR > 5.
\section{Conclusions}
\label{sec:conclude}
In this work, we calculated the Berry phase contribution to the scattering of quasiparticles from vortices in a d-wave superconductor. We simplified the Bogoliubov-de Gennes Hamiltonian by applying a singular gauge transformation. This transformation extracts the phase from the gap function and encodes it in the antiperiodic boundary conditions imposed on the wave function. Within the single vortex approximation, this antiperiodic boundary condition (Berry phase) manifests as a semi-infinite branch cut such that with each trip around the origin, the wave function changes sign. We neglected the superflow contribution and considered the scattering of quasiparticles due only to the presence of this antiperiodic boundary condition. We found the scattering cross section for this case to be divergent in the forward direction. In order to regularize the Berry phase effect we considered the two vortex problem (two Aharonov-Bohm half fluxes) which has a finite branch cut between its cores. To solve this two-center problem, we chose to work in elliptical coordinates. Elliptical coordinates provide an advantage in implementing the branch cut condition on the wave function spinor. We can turn on the Berry phase effect by simply imposing the boundary condition (see Eq.~(\ref{bound})) on the wave function spinor. We separated the (2+1)d Dirac equation in elliptical coordinates and found that the separated equations were Whittaker Hill equations (WHE). We solved the eigenvalue problem for the angular WHE. With the calculated eigenvalue as a parameter, we obtained the two linearly independent radial solutions with well defined asymptotic behavior. We developed a plane wave expansion for the incident quasiparticle current in terms of the separated solutions of the WHE. Using a partial wave analysis, we expressed the scattering amplitude and differential cross section in terms of phase shifts. We obtained the phase shifts by imposing the branch cut condition (Berry phase effect) on the full wave function spinor. We analyzed the scattering cross section due to the Berry phase effect for different separations of the vortex cores. We have also presented the variation of the total transport cross section as a function of the inter-vortex separation. We have also given a qualitative comparison of transport cross sections for the Berry phase and the superflow scattering mechanisms in Sec.~\ref{results}. Berry phase scattering of quasiparticles discussed here is not restricted to the case of d-wave superconductors. With some modifications of the incident plane wave, our problem becomes that of general relativistic scattering in two dimensions due to two Aharonov-Bohm half fluxes. In this work, we have neglected the superflow contribution. Single vortex scattering due to a circulating superflow is considered in a separate paper\cite{superflow1}. The problem that we have considered in both the manuscripts is a simplified version of a more complicated scattering process. We have made a series of approximations \cite{superflow1} to tackle the problem in its simplest form. This work should be treated as a first step forward to understand the complicated and important issue of scattering of quasiparticles from vortices. Deviating from the analytical setup, we can improve the model by considering the anisotropic Dirac spectrum, internodal scattering, and an even more rigorous description of plane waves. All these effects become important once we move away from the weak field limit. The linearized version of the BdG equation is limited to low energy quasiparticle excitations and one must use the full BdG equation and solve it numerically for higher energy cases. One way to include the above mentioned effects is to consider a sea of vortices in a lattice model. Such a vortex lattice calculation has been considered by Melikyan and Tesanovic \cite{tesanovic}. In the appendix of their paper, they have set up the two vortex scattering problem in elliptical coordinates. In this work, we have successfully obtained the exact scattering solutions to the two vortex problem in elliptical coordinates. The results of our calculations along with the vortex lattice calculations provide a greater insight into the bigger picture of quasiparticle scattering from vortices. To this end, our simplified model of double vortex scattering in an analytical framework is an important result. Our next step will be to consider both effects within the double vortex model by including the superflow that circulates around the vortices in the presence of the branch cut that lies between them.  We expect that this analysis, left for future research, will provide insight about not only the relative importance of the two contributions but also the interference between them.

\section{Acknowledgments}
S.G. would like to specially thank Alfred S. Goldhaber and Carl Bender for their valuable discussions and comments. We would also like to thank Sasha
Abanov, Patrick Lee, Zlatko Tesanovic, and Ashvin Vishwanath for very helpful discussions. This work was supported by the NSF under grant No. DMR-0605919. S.G. was also supported by the DOE under grant no. DE-FG02-09ER16052. M.K. was also supported by the NSF under grant No. DMR-0906866.

\appendix
\section{Bogoliubov-de Gennes Equation}
\label{sec:BdG}
The setup for this problem is described in detail in Ref.~\onlinecite{superflow1}. We summarize here. Consider the Bogoliubov-de Gennes (BdG) equation for a $d$-wave superconductor in the presence of a constant perpendicular magnetic field, ${\bf A}=\frac{1}{2}Hr\hat{\bf \phi}$, and with an order parameter that winds once about the origin, $\Delta({\bf r})=\Delta_{0}e^{i\phi}$:
\begin{equation}
H^{\prime} \Psi = E \Psi \;\;\;\;\;\;\;\;
H^{\prime} =
\left( \begin{array}{cc} \hat{H}_{e}^{\prime} & \hat{\Delta}^{\prime} \\
\hat{\Delta}^{\prime *} & -\hat{H}_{e}^{\prime *} \end{array} \right)
\label{eq:BdG}
\end{equation}
\begin{equation}
\hat{H}_{e}^{\prime} = \frac{1}{2m} \left( {\bf p}-\frac{e}{c}{\bf A}
\right)^{2} - E_{F}
\label{eq:He}
\end{equation}
\begin{equation}
\hat{\Delta}^{\prime} = \frac{1}{p_{F}^{2}}
\{\hat{p_{x}},\{\hat{p_{y}},\Delta({\bf r})\}\}
-\frac{i}{4p_{F}^{2}} \Delta({\bf r})
(\partial_{x} \partial_{y} \phi)
\label{eq:DeltaOp}
\end{equation}
Here ${\bf p}=-i\hbar {\bf \nabla}$, $\{a,b\}=(ab+ba)/2$, and $E$ is the quasiparticle energy. The form of the gap operator enforces the $d$-wave symmetry \cite{vafek,simon}. Upon circling an $hc/2e$ vortex, the quasiparticle acquires a Berry phase factor of (-1). This fact is encoded within the complex differential form of the gap operator, $\hat{\Delta}$. We simplify the Hamiltonian by effectively stripping the gap function, $\Delta({\bf r})$, of its phase. This is done by applying the singular gauge transformation
\begin{equation}
U = \left( \begin{array}{cc} e^{-i\phi/2} & 0 \\
0 & e^{i\phi/2} \end{array} \right)
\;\;\;\;\;\;\;\; \Phi({\bf r}) = U^{-1} \Psi({\bf r})
\;\;\;\;\;\;\;\; H = U^{-1} H^{\prime} U .
\label{eq:gaugetrans}
\end{equation}
In this gauge, known as the Anderson gauge,
\begin{equation}
H \Phi = E \Phi
\label{eq:transBdG}
\end{equation}
\begin{equation}
H = \tau_{3} \frac{v_{f}}{2p_{F}}
\left[ ({\bf p} + \tau_{3} {\bf P}_{s})^{2} - p_{F}^{2} \right]
+ \tau_{1} \frac{v_{2}}{2p_{F}}
\left[ 2 p_{x} p_{y} \right]
\label{eq:transH}
\end{equation}
where
\begin{equation}
{\bf P}_{s}({\bf r}) =
\frac{\hbar}{2} {\bf \nabla} \phi - \frac{e}{c} {\bf A} =
\frac{\hbar}{2} \left( \frac{1}{r} - \frac{r}{R^{2}} \right) \hat{\bf \phi}
\label{eq:ps}
\end{equation}
is the gauge invariant superfluid momentum (superflow), $v_{f}=p_{F}/m$, $v_{2}=\Delta_{0}/p_{F}$, and $R \equiv \sqrt{\hbar c / e H}$. In effect, the Berry phase contribution has been extracted from the Hamiltonian and encoded in the antiperiodic boundary conditions imposed on the wave function. While the original wave function was defined with periodic boundary conditions, $\Psi(r,\phi)=\Psi(r,\phi+2\pi)$, the transformed wave function is not single-valued and has antiperiodic boundary conditions, $\Phi(r,\phi)=-\Phi(r,\phi+2\pi)$. Hence, we have introduced a branch cut such that with each trip around the origin, the wave function changes sign. Note that the Berry phase effect is not a consequence of the choice of gauge. We have used the singular gauge transformation to extract the Berry phase contribution from the Hamiltonian and encode it in the boundary conditions of the wave functions. By definition, all observables, such as differential cross section or transport coefficients, are independent of this gauge choice.  For other problems, other gauge choices are optimal.  A nice discussion of this is provided for the case of vortex lattice by Franz and Tesanovic \cite{franztesano} and Vafek et al\cite{vafek-ashot}. We can further simplify our Hamiltonian by shifting the origin of momentum space to the location of one of the nodes. Shifting to node 1
\begin{equation}
p_{x} \rightarrow p_{F} + p_{x} \;\;\;\;\;\;\;\; p_{y} \rightarrow p_{y}
\label{eq:pshift}
\end{equation}
we find that
\begin{eqnarray}
H & = & H_{D} + H_{C}\label{eq:shiftedH}\\
H_{D} & = & v_{f} \left[ p_{x} \tau_{3} + \alpha p_{y} \tau_{1} + P_{sx} \right]\label{eq:HDirac}\\
H_{C} & = & \frac{v_{f}}{2p_{F}} \left[ (p^{2} + P_{s}^{2}) \tau_{3}
+ 2 {\bf P}_{s} \cdot {\bf p} + \alpha 2 p_{x} p_{y} \tau_{1} \right] \label{eq:Hcurve}
\end{eqnarray}
where $\alpha = v_{2}/v_{f}$ and we have used the fact that ${\bf P}_{s} = P_{s}(r) \hat{\bf \phi}$ to commute ${\bf p}$ with ${\bf P}_{s}$. Here $H$ is written as the sum of a linear (Dirac) Hamiltonian, $H_{D}$, and a quadratic (curvature) Hamiltonian, $H_{C}$. The second (curvature) term is smaller than the first by a factor of $E/E_{F}$. We will focus on the dominant term, $H_{D}$. In order to study the quasiparticle scattering from vortices, we must consider the nature of quasiparticle current in a $d$-wave superconductor. Since the incident and scattered currents will be considered in the far field where the quasiparticles are free, we wish to determine the quasiparticle current as a functional of $\Phi$ for $P_{s}=0$. Setting $P_{s}=0$ in Eq.~(\ref{eq:transH}) we find that the BdG Hamiltonian becomes
\begin{equation}
H = \left( \begin{array}{cc} \hat{H}_{e} & \hat{\Delta} \\
\hat{\Delta}^{*} & -\hat{H}_{e}^{*} \end{array} \right)
\;\;\;\;\;\;\;\; \hat{H}_{e} = - \frac{v_{f}}{2p_{F}} \nabla^{2} - E_{F}
\;\;\;\;\;\;\;\; \hat{\Delta} = - \frac{v_{2}}{2p_{F}}
2 \partial_{x} \partial_{y}
\label{eq:transHps0}
\end{equation}
Following Refs.~\onlinecite{superflow1} and \onlinecite{deGennes}, we can write down the quasiparticle current corresponding to the BdG Hamiltonian. Once again, it is convenient to shift the origin of momentum space to a nodal point. Shifting to node 1 yields
\begin{eqnarray}
{\bf j} & = & {\bf j}_{D} + {\bf j}_{C}
\label{eq:shiftedJ}\\
{\bf j}_{D} & = & v_{f} \Phi^{\dagger} (\tau_{3} \hat{\bf x}
+ \alpha \tau_{1} \hat{\bf y}) \Phi
\label{eq:JDirac}\\
{\bf j}_{C} & = & \frac{v_{f}}{p_{F}} \mbox{Im} \left[
\Phi^{\dagger} (\tau_{3} \hat{\bf x}
+ \alpha \tau_{1} \hat{\bf y}) \frac{\partial \Phi}{\partial x} +
\Phi^{\dagger} (\tau_{3} \hat{\bf y}
+ \alpha \tau_{1} \hat{\bf x}) \frac{\partial \Phi}{\partial y} \right]
\label{eq:Jcurve}
\end{eqnarray}
where $\alpha=v_{2}/v_{f}$.
In what follows, we will focus on the dominant term ${\bf j}_{D}$ corresponding to the Hamiltonian ${\bf H}_{D}$.

To proceed, we must obtain a general form for the scattering cross section. We consider a plane wave, with quasiparticle current in the incident direction, scattering off a vortex as a radial wave, with quasiparticle current in the scattered direction. If the incident momentum is ${\bf k} = (k,\theta)$ and the final momentum is ${\bf k}^{\prime} = (k,\phi)$, then the incident direction is the direction of the group velocity at momentum ${\bf k}$ and the scattered direction is the direction of the group velocity at momentum ${\bf k}^{\prime}$. For general, anisotropic nodes, the group velocity need not be parallel to the momentum. However, for the isotropic case that we consider
\begin{equation}
{\bf v}_{G}({\bf k}) = \frac{\partial E_{k}}{\partial {\bf k}}
= v_{f} \frac{\epsilon_{k}}{E_{k}} \hat{\bf x}
+ v_{2} \frac{\Delta_{k}}{E_{k}} \hat{\bf y}
= v_{f} \left( \cos\theta \hat{\bf x} + \sin\theta \hat{\bf y} \right)
= v_{f} \hat{\bf k}
\label{eq:groupvel}
\end{equation}
and the group velocity and momentum are parallel. Therefore, if $\Phi_{i}$ denotes the incident wave function and $\Phi_{s}$ denotes the scattered wave function, then we require
\begin{equation}
{\bf j}_{D}[\Phi_{i}] \sim
\left( \cos\theta \hat{\bf x} + \sin\theta \hat{\bf y} \right)
\sim \hat{\bf k} \;\;\;\;\;\;\;\;
{\bf j}_{D}[\Phi_{s}] \sim
\left( \cos\phi \hat{\bf x} + \sin\phi \hat{\bf y} \right)
\sim \hat{\bf k}^{\prime} \sim \hat{\bf r} .
\label{eq:JincJscat}
\end{equation}
Inspection of the form of the current functional,
${\bf j}_{D} = v_{f} \Phi^{\dagger}
(\tau_{3}\hat{\bf x} + \tau_{1}\hat{\bf y}) \Phi$
reveals that the appropriate incident plane wave is
\begin{equation}
\Phi_{i}({\bf r}) = e^{i{\bf k} \cdot {\bf r}}
\left( \begin{array}{c} \cos \frac{\theta}{2} \\
\sin \frac{\theta}{2} \end{array} \right)
\label{eq:Phii}
\end{equation}

Note that outside the vortex, quasiparticles are subject to neither an order parameter phase gradient nor a magnetic field.  Thus, the incident wave function is a plane wave.  This is consistent with the well-known results of Franz and Tesanovic\cite{franztesano} who showed that the low-energy quasiparticle states of a d-wave superconductor in the vortex state are Bloch waves of massless Dirac fermions rather than Landau Levels.  (For a discussion of the analysis that let to this important result, the reader is referred to Refs.~\onlinecite{gorkovSchrieffer, kopninvinokur, kopninbook, franztesano}.)

\begin{equation}
{\bf j}_{D}[\Phi_{i}] = v_{f} \left[
\left( \cos^{2} \frac{\theta}{2} - \sin^{2} \frac{\theta}{2} \right)
\hat{\bf x} + \left( 2 \sin \frac{\theta}{2} \cos \frac{\theta}{2} \right)
\hat{\bf y} \right] = v_{f} \hat{\bf k} .
\label{eq:jinc}
\end{equation}
Note also that this form solves the BdG equation, as it must in the absence of the vortex. The appropriate scattered radial wave is then given by
\begin{equation}
\Phi_{s}({\bf r}) = e^{i\frac{\phi}{2}} f(\phi,\theta) \frac{e^{ikr}}{\sqrt{r}}
\left( \begin{array}{c} \cos \frac{\phi}{2} \\
\sin \frac{\phi}{2} \end{array} \right)
\label{eq:Phis}
\end{equation}
\begin{equation}
{\bf j}_{D}[\Phi_{s}] = v_{f} \frac{|f|^{2}}{r} \left[
\left( \cos^{2} \frac{\phi}{2} - \sin^{2} \frac{\phi}{2} \right)
\hat{\bf x} + \left( 2 \sin \frac{\phi}{2} \cos \frac{\phi}{2} \right)
\hat{\bf y} \right] = v_{f} \frac{|f|^{2}}{r} \hat{\bf r} .
\label{eq:jscat}
\end{equation}
Here $f(\phi,\theta)$ is the scattering amplitude and the $e^{i\phi/2}$ prefactor has been added to make the
wave function single-valued.

\section{Separation of Variables for massless spin-1/2 2D Dirac equation in elliptical coordinates}
 \label{sov}

We present here the details of the steps (see Sec.~\ref{sec:Separation-of-Dirac}) leading to the separation of variables in elliptical coordinates of a two-dimensional Dirac equation for massless spin-1/2 fermions. This calculation is in the spirit of work done in Ref.~\onlinecite{1990-Villalba}. An alternate method can also be found in Ref.~\onlinecite{1982-cook}.
The 2-D Dirac equation is given by
\begin{equation}
[\gamma^{0}\partial_{t}+\gamma^{1}\partial_{x}+\gamma^{2}\partial_{y}]\psi=0\label{dirac}
\end{equation}
where the $\gamma$'s satisfy the following anticommutation relations
\begin{equation}
\{\gamma^{\alpha},\gamma^{\beta}\}=2g^{\alpha\beta}\label{clifford}
\end{equation}
with Minkowski metric given by $g=diag(1,-1,-1)$. In order to rewrite the Dirac equation in curved coordinates, let us introduce the following coordinate transformation
\begin{equation}
x=f(\mu,\nu), \quad y=g(\mu,\nu), \quad t=t\label{trans}.
\end{equation}
The reader should note that the transformations are kept general and the choice of elliptical coordinates will be made when required. The general transformations given by $f(\mu,\nu)$ and $g(\mu,\nu)$ must satisfy the condition that $f+ig$ is holomorphic or complex differentiable in the $u+iv$ plane which leads to the following Cauchy-Riemann equations,
\begin{equation}
f_{\mu}=g_{\nu},\quad g_{\mu}=-f_{\nu}\label{cauchy}
\end{equation}
Using Eq.~(\ref{trans}) and Eq.~(\ref{cauchy}) in Eq.~(\ref{dirac}), one can easily write the dirac equation in curved coordinates as
\begin{equation}
[\gamma^{0}\partial_{t}+\frac{\widetilde{\gamma}^{1}}{h}\partial_{\mu}+\frac{\widetilde{\gamma}^{2}}{h}\partial_{\nu}]\psi=0
\end{equation}
where
\begin{eqnarray}
\widetilde{\gamma}^{1} &=& \frac{1}{h}(f_{\mu}\gamma^{1}-f_{\nu}\gamma^{2})\\
\widetilde{\gamma}^{2} &=& \frac{1}{h}(f_{\nu}\gamma^{1}+f_{\mu}\gamma^{2})
\end{eqnarray}
with the Lame Metric given by $h=\sqrt{f_{\mu}^{2}+g_{\mu}^{2}}$. We introduce the following transformation matrices:
\begin{equation}
S=\frac{1}{\sqrt{h}}(e^{\frac{\phi}{2}\gamma^{1}\gamma^{2}}), \quad S^{-1}=\sqrt{h}(e^{-\frac{\phi}{2}\gamma^{1}\gamma^{2}})\label{S-matrix}
\end{equation}
with $\phi=\arctan(\frac{g_{\mu}}{f_{\mu}})$.
We use Eq.~(\ref{S-matrix}) in transforming the Dirac equation, That is we perform
\begin{equation}
S^{-1}[\gamma^{0}\partial_{t}+\frac{\widetilde{\gamma}^{1}}{h}\partial_{\mu}+\frac{\widetilde{\gamma}^{2}}{h}\partial_{\nu}]SS^{-1}\psi=0\label{sdirac}
\end{equation}
which results in
\begin{equation}
[\partial_{t}+\frac{\gamma^{0}\gamma^{1}}{h}\partial_{\mu}+\frac{\gamma^{0}\gamma^{2}}{h}\partial_{\nu}]\Phi=0\label{curved-dirac}
\end{equation}
where the transformed spinor satisfies
\begin{equation}
\Phi=S^{-1}\psi
\end{equation}
To separate the time variable we introduce the following operator definitions.
\begin{equation}
\widehat{k_{2}}\equiv\partial_{t}, \quad \widehat{k_{1}}\equiv(\frac{\gamma^{0}\gamma^{1}}{h}\partial_{\mu}+\frac{\gamma^{0}\gamma^{2}}{h}\partial_{\nu})
\end{equation}
The Hamiltonian Eq.~(\ref{curved-dirac}) will read
\begin{equation}
[\widehat{k_{2}}+\widehat{k_{1}}]\Phi=0
\end{equation}
with $\widehat{k_{1}}$ and $\widehat{k_{2}}$ satisfying the commutation relation
\begin{equation}
[\widehat{k_{2}},\widehat{k_{1}}]=0
\end{equation}
To separate the time variable, we introduce $k$ such that
\begin{equation}
\widehat{k_{2}}\Phi=-ik\Phi
\end{equation}
which immediately gives
\begin{equation}
\widehat{k_{1}}\Phi=ik\Phi\label{kiphi}
\end{equation}
Here, we make the following choice for a two-dimensional representation of the Dirac matrices,
\begin{equation}
\gamma^{0}=\tau_{2}, \quad \gamma^{1}=i\tau_{1} \quad, \gamma^{2}=-i\tau_{3}
\end{equation}
In this representation, Eq.~(\ref{kiphi}) reads
\begin{equation}
[\tau_{3}\partial_{\mu}+\tau_{1}\partial_{\nu}-ikh]\Phi=0\label{t-dirac}
\end{equation}
The presence of $h$ in Eq.~(\ref{t-dirac}) forbids us to write Eq.~(\ref{t-dirac}) as sum of two commuting differential operators. Therefore we will introduce a similarity transformation $T(\mu,\nu)$ acting on the Dirac operator
and the spinor.
\begin{equation}
T=e^{\beta}e^{i\alpha\tau_{2}}\label{T-tran}
\end{equation}
with
\begin{equation}
\alpha_{\mu}=-\beta_{\nu}, \quad \alpha_{\nu}=\beta_{\mu}\label{cauchy-alpha}
\end{equation}
To be more explicit, we do the T-transformation on Eq.~(\ref{t-dirac}) in the following way
\begin{equation}
T[\tau_{3}\partial_{\mu}+\tau_{1}\partial_{\nu}-ikh]TT^{-1}\Phi=0
\end{equation}
which after some algebra gives,
\begin{equation}
[\tau_{3}\partial_{\mu}+\tau_{1}\partial_{\nu}-ikhe^{i2\alpha\tau_{2}}]Y=0\label{dirac-Y}
\end{equation}
with
\begin{equation}
Y=T^{-1}\Phi
\end{equation}
It should be noted that because $\alpha(\mu,\nu)$ in Eq.~(\ref{T-tran}) is arbitrary we choose it to be of the following structure,
\begin{equation}
e^{i2\alpha\tau_{2}}=\frac{[a(\mu)+ib(\nu)\tau_{2}]}{h}
\end{equation}
This specific form of the transformation matrix T cancels the factor h (Lame metric) in the Dirac equation which mixes the $\mu$ and $\nu$ variables.
It is trivial to check that $a(\mu)^{2}+b(\nu)^{2}=h^{2}$ and the structure for $\beta(\mu,\nu)$ can be obtained using Eq.~(\ref{cauchy-alpha}).
At this point we move to elliptical coordinates with Eq.~(\ref{trans}) taking the form
\begin{eqnarray}
\label{elliptic1}
x=f(\mu,\nu)=R\,\cosh\mu\,\cos\nu \\\label{elliptic2}
y=g(\mu,\nu)=R\,\sinh\mu\,\sin\nu
\end{eqnarray}
From (\ref{elliptic1}, \ref{elliptic2}) we get
\begin{equation}
h^{2}=R^{2}\sinh^{2}\mu+R^{2}\sin^{2}\nu\label{h2}
\end{equation}
Using above (\ref{h2}) along with the fact that $a(\mu)^{2}+b(\nu)^{2}=h^{2}$ gives us by comparison the following
\begin{equation}
a(\mu)=R\,\sinh\mu,\quad b(\nu)=R\,\sin\nu\
\end{equation}
Now we introduce the following operators
\begin{eqnarray}
\label{L1L2Y}
\widehat{L_{1}}=\tau_{3}\partial_{\mu}-ika(\mu)\\\label{L1L2Ya}
\widehat{L_{2}}=\tau_{3}\partial_{\nu}-ikb(\nu)\\\label{L1L2Yb}
Y=[\widehat{L_{2}}+i\tau_{2}\widehat{L_{1}}]Z\label{YZ}
\end{eqnarray}
Using Eqs.~(\ref{L1L2Y}), (\ref{L1L2Ya}), and (\ref{L1L2Yb}) we can finally express the Dirac equation Eq.~(\ref{dirac-Y}) as
\begin{equation}
[(\partial_{\mu}^{2}-ik\tau_{3}a_{\mu}+k^{2}a^{2})+(\partial_{\nu}^{2}-ik\tau_{3}b_{\nu}+k^{2}b^{2})]Z=0
\end{equation}
Defining
\begin{equation}
Z=\left(\begin{array}{c}
\alpha(\mu)A(\nu)\\
\beta(\mu)B(\nu)\end{array}\right)
\end{equation}
and introducing a separation constant $\lambda$ gives us four 2nd order ordinary differential equations.
\begin{eqnarray}
\label{ode1}
(\partial_{\mu}^{2}-ika_{\mu}+k^{2}a^{2}-\lambda^{2})\alpha(\mu)=0\\\label{ode2}
(\partial_{\mu}^{2}+ika_{\mu}+k^{2}a^{2}-\lambda^{2})\beta(\mu)=0\\\label{ode3}
(\partial_{\nu}^{2}-ikb_{\nu}+k^{2}b^{2}+\lambda^{2})A(\nu)=0\\\label{ode4}
(\partial_{\nu}^{2}+ikb_{\nu}+k^{2}b^{2}+\lambda^{2})B(\nu)=0
\end{eqnarray}
After putting back $a=R\sinh\mu$, $a_{\mu}=R\cosh\mu$, $b=R\sin\nu$, and $b_{\nu}=R\cos\nu$ we obtain
\begin{eqnarray}
(\partial_{\nu}^{2}-ikR\cos\nu+k^{2}R^{2}\sin^{2}\nu+\lambda^{2})A(\nu) & = & 0\\
\ (\partial_{\nu}^{2}+ikR\cos\nu+k^{2}R^{2}\sin^{2}\nu+\lambda^{2})B(\nu) & = & 0\\
(\partial_{\mu}^{2}-ikR\cosh\mu+k^{2}R^{2}\sinh^{2}\mu-\lambda^{2})\alpha(\mu) & = & 0\\
(\partial_{\mu}^{2}+ikR\cosh\mu+k^{2}R^{2}\sinh^{2}\mu-\lambda^{2})\beta(\mu) & = & 0\label{eq:whe1}\end{eqnarray}
which are equivalent to 4 coupled first order equations that connect the upper and lower components of the wave function spinor
\begin{eqnarray}
(\partial_{\nu}-ikR\sin\nu)A(\nu) & = & i\lambda B(\nu)\\
(\partial_{\nu}+ikR\sin\nu)B(\nu) & = & i\lambda A(\nu)\\
(\partial_{\mu}-ikR\sinh\mu)\alpha(\mu) & = & \lambda\beta(\mu)\\
(\partial_{\mu}+ikR\sinh\mu)\beta(\mu) & = & \lambda\alpha(\mu)\label{eq:couple1}\end{eqnarray}
Thus we have reduced the 2D massless Dirac equation in elliptical coordinates to a problem of four decoupled ordinary differential equations. These separated radial and angular equations are known as the Whittaker Hill equations (WHE).

The explicit form of the transformation matrices can be evaluated. Using Eq.~(\ref{YZ}) we have
\begin{equation}
Y=\left(\begin{array}{c}
\alpha(\mu)B(\nu)\\
i\beta(\mu)A(\nu)\end{array}\right)\end{equation}
We write below a more transparent and dimensionless form for $S$ (Eq.~(\ref{S-matrix})) and $T$ (Eq.~(\ref{T-tran}))
\begin{eqnarray}
S & = & \frac{\sqrt{R}}{\sqrt{h}}\left(\begin{array}{cc}
\cos\frac{\phi}{2} & -\sin\frac{\phi}{2}\\
\sin\frac{\phi}{2} & \cos\frac{\phi}{2}\end{array}\right)\\
T & = & e^{\beta}\left(\begin{array}{cc}
\cos\alpha & \sin\alpha\\
-\sin\alpha & \cos\alpha\end{array}\right)
\end{eqnarray}
which immediately gives $ST$ as
\begin{equation}
ST=\sqrt{R}\frac{e^{\beta}}{\sqrt{h}}\left(\begin{array}{cc}
\cos(\alpha-\phi/2) & \sin(\alpha-\phi/2)\\
-\sin(\alpha-\phi/2) & \cos(\alpha-\phi/2)\end{array}\right)
\label{STT}
\end{equation}
In elliptical coordinates, $\alpha$, $\beta$, and $\phi$ take the following form,
\begin{eqnarray}
\alpha & = & \frac{1}{2}\tan^{-1}(\frac{\sin\,\nu}{\sinh\,\mu})\\
\beta & = & -\frac{1}{2}\tanh^{-1}(\frac{\cos\,\nu}{\cosh\,\mu})\\
\phi & = & \tan^{-1}(\coth\,\mu\, \tan\,\nu)\\
\end{eqnarray}
and we can rewrite $\beta$ using the log formula for $\tanh^{-1}$.
\begin{eqnarray}
\beta & = & \ln(\frac{\cosh\,\mu+\cos\,\nu}{\cosh\,\mu-\cos\,\nu})^{-1/4}\label{newbeta}\\
\frac{e^{\beta}}{\sqrt{h}} & = & \frac{1}{\sqrt{(\cosh\,\mu+\cos\,\nu)}}\label{expbeta}\\
\end{eqnarray}
After using some trigonometric identities, these expressions for $\alpha$ and $\phi$ satisfy the following relations
\begin{eqnarray}
\alpha-\frac{\phi}{2} & = & \frac{\tan^{-1}(\frac{\sin\,\nu}{\sinh\,\mu})-\tan^{-1}(\coth\,\mu\,\tan\,\nu)}{2}\label{transangle}\\
2(\alpha-\frac{\phi}{2}) & = & \tan^{-1}(\frac{-\sin\,\nu\,\sinh\,\mu}{\cos\,\nu\,\cosh\,\mu+1})\end{eqnarray}
To evaluate ST we need to calculate its components in terms of elliptical coordinate variables. Tangent, sine and cosine of the double angle can be evaluated as
\begin{eqnarray*}
\tan(2(\alpha-\frac{\phi}{2})) & = & \frac{-\sin\,\nu\,\sinh\,\mu}{\cos\,\nu\,\cosh\,\mu+1}\\
\sin(2(\alpha-\frac{\phi}{2})) & = & \frac{-\sin\,\nu\,\sinh\,\mu}{\cos\,\nu+\cosh\,\mu}\\
\cos(2(\alpha-\frac{\phi}{2})) & = & \frac{\cos\,\nu\cosh\,\mu+1}{\cos\,\nu+\cosh\,\mu}\\
\end{eqnarray*}
Using the half angle formulae we can finally evaluate the components of the transformation matrix in terms of the coordinate variables. While reducing the components to half angle there is arbitrariness in the choice of the sign of the trigonometric functions in each quadrant.
We make a choice of signs such that our final transformation matrix is 4$\pi$ periodic in angular coordinate $\nu$. We could as well make a choice which would result in 2$\pi$ periodicity of transformation matrix. Any choice can be compensated by choosing the periodicity of the angular solutions accordingly. For example, we choose 4$\pi$ periodic eigenstates for the angular WHE (see Sec.~\ref{sec:Solutions-to-Whittakker-Hill's}).
\begin{eqnarray}
\sin(\alpha-\frac{\phi}{2}) & = & -\frac{\sin\,\frac{\nu}{2}\,\sinh\,\frac{\mu}{2}}{\sqrt{\cos\,\nu+\cosh\,\mu}}\label{sinap}\\
\cos(\alpha-\frac{\phi}{2}) & = & \frac{\cos\,\frac{\nu}{2}\cosh\,\frac{\mu}{2}}{\sqrt{\cos\,\nu+\cosh\,\mu}}\label{cosap}\\
\end{eqnarray}
Using Eqs.~(\ref{STT}, (\ref{expbeta}), (\ref{sinap}), and (\ref{cosap}) along with the fact that $\psi=ST\,Y$, we have
\[\psi=\frac{1}{(\cosh\,\mu+\cos\,\nu)}\left(\begin{array}{cc}
\cos\,\frac{\nu}{2}\cosh\,\frac{\mu}{2} & -\sin\,\frac{\nu}{2}\sinh\,\frac{\mu}{2}\\
\sin\,\frac{\nu}{2}\sinh\,\frac{\mu}{2} & \cos\,\frac{\nu}{2}\cosh\,\frac{\mu}{2}\end{array}\right)\left(\begin{array}{c}
\alpha(\mu)B(\nu)\\
i\beta(\mu)A(\nu)\end{array}\right)\]

\section{Asymptotic form for radial solutions}
\label{sec:Appendix--Asymptotic-form}

To study the scattering cross section, we have to calculate the phase shifts of the scattered wave function at large radial distances where the detector is placed. Hence, the asymptotic form of the radial solutions is an important piece of information in setting up the scattering cross section. In this section we give the asymptotic form of the first and second radial solutions to WHE. In the Sec.~\ref{sub:Radia} we expand radial solutions in terms of confluent hypergeometric functions (CHF) (see Eqs.~(\ref{eq:evenhyp}) and (\ref{eq:oddhyp})). We now write the asymptotic
form of these solutions. As $\mu\rightarrow\infty$ in the series of CHF, only the leading term in the sum contributes
\begin{eqnarray}
Jo_{m}(\mu) & \approx & e^{i kR\cosh\mu}\ \cosh\frac{\mu}{2}\ M(\frac{1}{2},2,-4ikR\cosh^{2}\frac{\mu}{2})\\
Je_{m}(\mu) & \approx & e^{i kR\cosh\mu}\ \sinh\frac{\mu}{2}\ M(\frac{1}{2},2,-4ikR\cosh^{2}\frac{\mu}{2})\end{eqnarray}
The asymptotic form of the CHF is well known \cite{grad} and is given as
\begin{equation}
M(\frac{1}{2},2,-4ikR\cosh^{2}\frac{\mu}{2})\approx\frac{1}{\Gamma(\frac{1}{2})}\sqrt{\frac{1}{4ikR\cosh^{2}\frac{\mu}{2}}}\end{equation}
At large distances, we have $\cosh\mu\sim\sinh\mu$$\sim$$e^{\mu}=\frac{2r}{R}$, the elliptic coordinate $\nu$ is reduced to the ordinary polar angle $\phi$, and therefore the full asymptotic form of the solutions become\begin{eqnarray}
Jo_{m}(r) & \approx & \sqrt{\frac{1}{\pi kR}}e^{i(kr-\frac{\pi}{4})}\ \label{eq:largejo}\\
Je_{m}(r) & \approx & \sqrt{\frac{1}{\pi kR}}e^{i(kr-\frac{\pi}{4})}\ \label{eq:largeje}\end{eqnarray}
From inspection of the large $\mu$ behavior of the second solutions we find the asymptotic form to be
\begin{eqnarray}
Fey_{m}(r) & \approx & \sqrt{\frac{1}{\pi kR}}e^{i(kr+\frac{\pi}{4})}\label{eq:largefey}\\
Gey_{m}(r) & \approx & \sqrt{\frac{1}{\pi kR}}e^{i(kr+\frac{\pi}{4})}\ \ \label{eq:largegey}\end{eqnarray}
Similarly we can also determine the asymptotic forms for the lower component radial solutions
\begin{eqnarray}
Jo_{m}^{'}(r) & \approx & \sqrt{\frac{1}{\pi kR}}e^{-im\pi}e^{-i(kr+\frac{\pi}{4})}\label{eq:largejop}\\
Je_{m}^{'}(r) & \approx & \sqrt{\frac{1}{\pi kR}}e^{-im\pi}e^{-i(kr+\frac{\pi}{4})}\ \label{eq:largejep}\\
Fey_{m}^{'}(r) & \approx & \sqrt{\frac{1}{\pi kR}}e^{-im\pi}e^{-i(kr-\frac{\pi}{4})}\ \label{eq:largefeyp}\\
Gey_{m}^{'}(r) & \approx & \sqrt{\frac{1}{\pi kR}}e^{-im\pi}e^{-i(kr-\frac{\pi}{4})}\ \label{eq:largegeyp}\end{eqnarray}

\section{Plane wave expansion coefficients}
\label{sec:Appendix--Plane-wave}

In this section we calculate the plane wave expansion coefficients $n_{m}^{e}$ and $n_{m}^{o}$ appearing in Eq.~(\ref{eq:planeexp}).
\begin{equation}
e^{i\,\vec{k}\cdot\vec{r}}\left(\begin{array}{c}
\cos\frac{\theta}{2}\\
\sin\frac{\theta}{2}\end{array}\right)=(ST)\left\{ \sum_{m}n_{m}^{e}Be_{m}(\theta)\left(\begin{array}{c}
Je_{m}Be_{m}\\
i\, Je_{m}^{'}Ae_{m}\end{array}\right)+\sum_{m}n_{m}^{o}Bo_{m}(\theta)\left(\begin{array}{c}
Jo_{m}Bo_{m}\\
i\, Jo_{m}^{'}Ao_{m}\end{array}\right)\right\} \label{eq:planeexpapp}\end{equation}
In elliptical coordinates a unidirectional plane wave is described in terms of the incident direction $\theta$ in the following way,
\begin{eqnarray}
\vec{k} & = & (\cos\theta\hat{x}+\sin\theta\hat{y}),\ \vec{r}=R(\cosh\mu\cos\nu\ \hat{x}+\sinh\mu\sin\nu\ \hat{y})\\
\vec{k}\cdot\vec{r} & = & kR(\cosh\mu\cos\nu\cos\theta+\sinh\mu\sin\nu\sin\theta)\end{eqnarray}
Multiplying both sides by the transformation matrix $(ST)^{-1}$,
\begin{eqnarray*}
e^{i\,\vec{k}\cdot\vec{r}}(ST)^{-1}\left(\begin{array}{c}
\cos\frac{\theta}{2}\\
\sin\frac{\theta}{2}\end{array}\right) & = & \left\{ \sum_{m}n_{m}^{e}Be_{m}(\theta)\left(\begin{array}{c}
Je_{m}Be_{m}\\
i\, Je_{m}^{'}Ae_{m}\end{array}\right)+\sum_{m}n_{m}^{o}Bo_{m}(\theta)\left(\begin{array}{c}
Jo_{m}Bo_{m}\\
i\, Jo_{m}^{'}Ao_{m}\end{array}\right)\right\} \\
e^{i\,\vec{k}\cdot\vec{r}}\left(\begin{array}{c}
\cos\frac{\nu}{2}\cosh\frac{\mu}{2}\cos\frac{\theta}{2}+\sin\frac{\nu}{2}\sinh\frac{\mu}{2}\sin\frac{\theta}{2}\\
\cos\frac{\nu}{2}\cosh\frac{\mu}{2}\sin\frac{\theta}{2}-\sin\frac{\nu}{2}\sinh\frac{\mu}{2}\cos\frac{\theta}{2}\end{array}\right) & = & \left\{ \sum_{m}n_{m}^{e}Be_{m}(\theta)\left(\begin{array}{c}
Je_{m}Be_{m}\\
i\, Je_{m}^{'}Ae_{m}\end{array}\right)+\sum_{m}n_{m}^{o}Bo_{m}(\theta)\left(\begin{array}{c}
Jo_{m}Bo_{m}\\
i\, Jo_{m}^{'}Ao_{m}\end{array}\right)\right\} \end{eqnarray*}
For brevity of notation, we write\begin{equation}
\left(\begin{array}{c}
\cos\frac{\nu}{2}\cosh\frac{\mu}{2}\cos\frac{\theta}{2}+\sin\frac{\nu}{2}\sinh\frac{\mu}{2}\sin\frac{\theta}{2}\\
\cos\frac{\nu}{2}\cosh\frac{\mu}{2}\sin\frac{\theta}{2}-\sin\frac{\nu}{2}\sinh\frac{\mu}{2}\cos\frac{\theta}{2}\end{array}\right)\equiv\left(\begin{array}{c}
f_{up}(\mu,\nu,\theta)\\
f_{d}(\mu,\nu,\theta)\end{array}\right)\end{equation}
We also have the following bi-orthogonal properties for the angular Whittaker Hill functions.
\begin{eqnarray}
\intop_{0}^{2\pi}Ao_{s}^{*}Be_{m}d\nu & = & k_{m}\delta_{ms}\ ,\ \ \intop_{0}^{2\pi}Ae_{s}^{*}Bo_{m}d\nu=k_{m}^{*}\delta_{ms}\\
\intop_{0}^{2\pi}Ae_{s}^{*}Be_{m}d\nu & = & 0,\ \ \intop_{0}^{2\pi}Ao_{s}^{*}Bo_{m}d\nu=0\end{eqnarray}
where $k_{m}$ is the normalization constant.
Operating on both sides with row vector $\left(\begin{array}{cc}
Ao_{s}^{*}(\theta)Ao_{s}^{*}(\nu), & Ao_{s}^{*}(\theta)Bo_{s}^{*}(\nu)\end{array}\right)$, integrating over $\nu$ and $\theta$ from $0$ to $2\pi$, and applying the bi-orthogonality relations, we get
\[
\intop_{0}^{2\pi}\intop_{0}^{2\pi}e^{i\, kR(\cosh\mu\cos\nu\cos\theta+\sinh\mu\sin\nu\sin\theta)}Ao_{s}^{*}(\theta)\left(\begin{array}{cc}
Ao_{s}^{*}(\nu)f_{up}(\mu,\nu,\theta)+ & Bo_{s}^{*}(\nu)f_{d}(\mu,\nu,\theta)\end{array}\right)d\nu\ d\theta=\left\{ n_{s}^{e}\ k_{s}^{2}(Je_{s}+iJe_{s}^{'})\right\} \]
Putting $\mu=\mu_{0}$ in the above expression yields
\begin{equation}
n_{s}^{e}=\frac{\intop_{0}^{2\pi}\intop_{0}^{2\pi}e^{i\, kR(\cosh\mu_{0}\cos\nu\cos\theta+\sinh\mu_{0}\sin\nu\sin\theta)}Ao_{s}^{*}(\theta)\left(\begin{array}{cc}
Ao_{s}^{*}(\nu)f_{up}(\mu_{0},\nu,\theta)+ & Bo_{s}^{*}(\nu)f_{d}(\mu_{0},\nu,\theta)\end{array}\right)d\nu\ d\theta}{k_{s}^{2}(Je_{s}(\mu_{0})+iJe_{s}^{'}(\mu_{0}))}\label{eq:pwexpeven}\end{equation}

Similarly, we can calculate $n_{s}^{o}$ by operating on both sides with row vector $\left(\begin{array}{cc}
Ae_{s}^{*}(\theta)Ae_{s}^{*}(\nu), & Ae_{s}^{*}(\theta) Be_{s}^{*}(\nu)\end{array}\right)$ \begin{equation}
n_{s}^{o}=\frac{\intop_{0}^{2\pi}\intop_{0}^{2\pi}e^{i\, kR(\cosh\mu_{0}\cos\nu\cos\theta+\sinh\mu_{0}\sin\nu\sin\theta)}Ae_{s}^{*}(\theta)\left(\begin{array}{cc}
Ae_{s}^{*}(\nu)f_{up}(\mu_{0},\nu,\theta)+ & Be_{s}^{*}(\nu)f_{d}(\mu_{0},\nu,\theta)\end{array}\right)d\nu d\theta}{(k_{s}^{*})^{2}(Jo_{s}(\mu_{0})+iJo_{s}^{'}(\mu_{0}))}\label{eq:pwexpodd}\end{equation}

We have checked that $n_{s}^{e}$ and $n_{s}^{o}$ are independent of the value of $\mu_{0}$.

\bibliography{apssamp}
\end{document}